\documentclass[aps,prd,twocolumn,preprintnumbers,
superscriptaddress,showpacs,floatfix]{revtex4}

\usepackage{bm}
\usepackage{epsfig}
\usepackage{graphics}
\usepackage{amsmath}

\begin{document}

\preprint{UT-STPD-13/01}

\def\beq{\begin{equation}}
\def\eeq{\end{equation}}
\def\bea{\begin{eqnarray}}
\def\eea{\end{eqnarray}}
\newcommand{\eem}{\end{matrix}}
\newcommand{\bem}{\begin{matrix}}
\newcommand{\beqs}{\begin{subequations}}
\newcommand{\eeqs}{\end{subequations}}
\newcommand{\ftn}{\footnotesize}
\newcommand{\nsz}{\normalsize}
\newcommand{\ssz}{\scriptsize}

\newcommand{\dPp}{\ensuremath{\delta\Phi_+}}
\newcommand{\dPm}{\ensuremath{\delta\Phi_-}}
\newcommand{\dT}{\ensuremath{\delta T}}
\newcommand{\dS}{\ensuremath{\delta S}}
\newcommand{\dTb}{\ensuremath{\delta\bar T}}
\newcommand{\dP}{\ensuremath{\delta\Phi}}
\newcommand{\dPb}{\ensuremath{\delta\bar\Phi}}
\newcommand{\mqn}{\ensuremath{m_{\rm I-}}}
\newcommand{\Gqn}{\ensuremath{\Gm[{\rm I}-]}}
\newcommand{\Gqp}{\ensuremath{\Gm[{\rm I}+]}}
\newcommand{\MeV}{\ensuremath{\rm MeV}}

\newcommand{\vp}{\ensuremath{v_\Phi}}
\newcommand{\vT}{\ensuremath{v_T}}

\newcommand{\msp}{\ensuremath{m_{\rm I+}}}
\newcommand{\msm}{\ensuremath{m_{\rm I-}}}

\newcommand{\Eref}[1]{Eq.~(\ref{#1})}
\newcommand{\Sref}[1]{Sec.~\ref{#1}}
\newcommand{\Fref}[1]{Fig.~\ref{#1}}
\newcommand{\Tref}[1]{Table~\ref{#1}}
\newcommand{\cref}[1]{Ref.~\cite{#1}}
\newcommand\eqs[2]{Eqs.~(\ref{#1}) and (\ref{#2})}
\newcommand{\sEref}[2]{Eq.~(\ref{#1}{\small\sf {#2}})}

\newcommand\vev[1]{\langle {#1} \rangle}
\def\openep{\leavevmode\hbox{
{\boldmath $\varepsilon$}}}
\def\llgm{\left\lgroup}
\def\rrgm{\right\rgroup}
\newcommand\matt[4]{\mbox{$\llgm\bem #1 &#2 \cr #3& #4\eem\rrgm$}}

\def\lf{\left(}
\def\rg{\right)}
\def\lfa{\left|}
\def\rga{\right|}
\newcommand{\etal}{{\it et al.}}
\newcommand{\GeV}{{\mbox{\rm GeV}}}
\newcommand{\tnb}{{\ensuremath{\tan\beta}}}
\newcommand{\Mgut}{\ensuremath{M_{\rm GUT}}}
\newcommand{\Ggut}{\ensuremath{G_{\rm LR}}}
\newcommand{\Gsm}{\ensuremath{G_{\rm SM}}}
\newcommand{\ldt}{\ensuremath{\lambda_{T}}}
\newcommand{\lds}{\ensuremath{\lambda_{T}'}}
\newcommand{\ld}{\ensuremath{\lambda}}
\newcommand{\TeV}{\ensuremath{\rm TeV}}
\newcommand{\eV}{{\mbox{\rm eV}}}

\def\ve{\varepsilon}
\def\aal{{\bar\alpha}}
\def\bbet{{\bar\beta}}
\def\al{{\alpha}}
\def\bt{{\beta}}
\def\to{\rightarrow}
\def\dg{\dagger}
\def\llgm{\left\lgroup}
\def\rrgm{\right\rgroup}
\def\lf{\left(}
\def\rg{\right)}

\newcommand{\Gr}{\ensuremath{\widetilde{G}}}
\newcommand{\Yb}{\ensuremath{Y_{B}}}
\newcommand{\Yg}{\ensuremath{Y_{\Gr}}}
\newcommand{\Vhi}{\ensuremath{V_{\rm HI}}}
\newcommand{\Vhios}{\ensuremath{V_{\rm HI0}^{\rm SG}}}
\newcommand{\Hhi}{\ensuremath{H_{\rm HI}}}
\newcommand{\Ohi}{\ensuremath{\Omega}}
\newcommand{\Omg}{\ensuremath{\Omega}}
\newcommand{\Khi}{\ensuremath{K}}
\newcommand{\Vhio}{\ensuremath{V_{\rm HI0}}}
\newcommand{\Vd}{\ensuremath{V_{\rm D}}}
\newcommand{\cks}{{\mbox{\sl a}}}
\newcommand{\ckss}{{\mbox{\ssz\sl a}}}
\newcommand{\mP}{\ensuremath{m_{\rm P}}}
\newcommand{\Mpq}{\ensuremath{M}}
\newcommand{\mpq}{\ensuremath{m_{\rm PS}}}

\def\FHI{FHI~}
\newcommand{\Tr}{\mbox{\sf Tr}}
\newcommand{\diag}{\ensuremath{{\sf diag}}}
\newcommand{\im}{\ensuremath{{\sf Im}}}
\newcommand{\br}{\ensuremath{{\sf Br}}}
\newcommand{\tr}{{\mbox{\sf\ssz T}}}

\newcommand{\Ms}{\ensuremath{M_{\rm S}}}
\newcommand{\la}{\ensuremath{\lambda_a}}
\newcommand{\lm}{\ensuremath{\lambda_\mu}}
\newcommand{\kpt}{\ensuremath{\kappa_T}}
\newcommand{\mt}{\ensuremath{M_T}}
\newcommand{\lnu}{\ensuremath{\lambda_{\nu}}}
\newcommand\lvc[3]{\ensuremath{\varepsilon_{{#1}{#2}{#3}}}}
\def\openep{\leavevmode\hbox{\normalsize{\boldmath $\varepsilon$}}}
\def\openone{\leavevmode\hbox{\small1\kern-3.8pt\normalsize1}}

\newcommand{\ca}{\ensuremath{{\rm a}}}

\newcommand{\Gsn}{\ensuremath{\Gamma_{\rm I}}}
\newcommand{\GNsn[1]}{\ensuremath{\Gamma_{{\rm I}{#1}\nu^c}}}
\newcommand{\Ghsn}{\ensuremath{\Gamma_{{\rm I}y}}}
\newcommand{\Gm[1]}{\ensuremath{\Gamma_{#1}}}
\newcommand{\msn}{\ensuremath{m_{\rm I}}}

\newcommand{\hef}{\ensuremath{\ld_{1\nu^c}}}

\newcommand{\ef}{{\ensuremath\mbox{\slshape\sffamily N}}}
\newcommand{\ent}{{\ensuremath\mbox{\slshape\sffamily s}}}

\newcommand{\hd}{{\ensuremath{H_1}}}
\newcommand{\hu}{{\ensuremath{H_2}}}

\newcommand{\chir}{\ensuremath{c_{\rm HI}}}
\newcommand{\Nhi}{\ensuremath{N_{\rm HI*}}}

\newcommand{\ck}{\ensuremath{c_{2K}}}
\newcommand{\ckk}{\ensuremath{c_{4K}}}
\newcommand{\ckx}{\ensuremath{c_{6K}}}
\newcommand{\ckh}{\ensuremath{c_{8K}}}
\newcommand{\ckt}{\ensuremath{c_{10K}}}
\newcommand{\kp}{\ensuremath{\kappa}}
\newcommand{\ks}{\ensuremath{k_{4S}}}
\newcommand{\kss}{\ensuremath{k_{6S}}}
\newcommand{\ksss}{\ensuremath{k_{8S}}}
\newcommand{\ksh}{\ensuremath{k_{8S}}}
\newcommand{\kst}{\ensuremath{k_{10S}}}
\newcommand{\ksv}{\ensuremath{k_{12S}}}
\newcommand{\vg}{\ensuremath{v_{_G}}}

\newcommand{\Dex}{\ensuremath{\Delta_{\rm m*}}}

\newcommand{\phh}{\ensuremath{\Phi}}
\newcommand{\bph}{\ensuremath{\bar \Phi}}

\newcommand{\what}{\ensuremath{\widehat}}
\newcommand{\sgm}{\ensuremath{\sigma}}

\newcommand{\mps}{\ensuremath{m_{+}}}
\newcommand{\mng}{\ensuremath{m_{-}}}
\newcommand{\Dp}{\ensuremath{D_{+}}}
\newcommand{\Dn}{\ensuremath{D_{-}}}

\def\Ka{K\"{a}hler potential}
\def\Kap{K\"{a}hler potential}
\def\p{|S|}

\newcommand{\ns}{\ensuremath{n_{\rm s}}}
\newcommand{\as}{\ensuremath{\alpha_{\rm s}}}

\newcommand{\rcc}{\ensuremath{\mathcal{R}}}
\newcommand{\rce}{\ensuremath{\widehat{\mathcal{R}}}}
\newcommand{\sni}{\ensuremath{\nu^c_i}}
\newcommand{\sn}{\ensuremath{\widetilde N}}
\newcommand{\ssni}{\ensuremath{\tilde \nu_i^c}}
\newcommand{\rhn}{\ensuremath{\nu^c}}
\newcommand{\rhni}{\ensuremath{\nu^c_i}}
\newcommand{\wrhn[1]}{\ensuremath{\nu^c_{#1}}}
\newcommand{\mrh[1]}{\ensuremath{M_{{#1}\nu^c}}}
\newcommand{\mD[1]}{\ensuremath{m_{#1\rm D}}}
\newcommand{\mn[1]}{\ensuremath{m_{#1\nu}}}
\newcommand{\mntau}{\ensuremath{m_{\nu_\tau}}}
\newcommand{\mnmu}{\ensuremath{m_{\nu_\mu}}}
\newcommand{\cth}{\ensuremath{c_\vartheta}}
\newcommand{\sth}{\ensuremath{s_\vartheta}}

\newcommand{\Trh}{\ensuremath{T_{\rm rh}}}
\newcommand{\sg}{\ensuremath{h}}
\newcommand{\sgc}{\ensuremath{\sg_{\rm c}}}
\newcommand{\sgf}{\ensuremath{\sg_{\rm f}}}

\newcommand{\snH}{\ensuremath{\Phi}}
\newcommand{\snHb}{\ensuremath{\bar\Phi}}

\newcommand{\hh}{{\ensuremath{
I{\kern-2.6pt h}}}}
\newcommand{\bhh}{{\ensuremath{\bar{
I{\kern-2.6pt h}}}}}

\renewcommand\mtt[4]{\mbox{ $\llgm\bem #1 &#2 \cr #3&
#4\eem\rrgm$}}

\renewcommand{\arraystretch}{1.4}

\newcommand\lin[2]{\mbox{$\llgm\bem #1& #2\eem\rrgm$}}

\newcommand\stl[2]{\mbox{$\llgm\bem #1\cr #2\eem\rrgm$}}

\newcommand\mtn[9]{\ensuremath{\llgm\bem #1&#2&#3\cr #4&#5&#6 \cr #7&#8&#9\eem\rrgm}}

\renewenvironment{subequations}{%
\refstepcounter{equation}%
\setcounter{parentequation}{\value{equation}}%
  \setcounter{equation}{0}
  \def\theequation{\theparentequation{\sffamily\alph{equation}}}%
  \ignorespaces
}{%
  \setcounter{equation}{\value{parentequation}}%
  \ignorespacesafterend
}

\title{Inflation, Leptogenesis, and Yukawa Quasi-Unification \\ within a
Supersymmetric Left-Right Model}

\author{R. Armillis}
\email{roberta.armillis@epfl.ch} \affiliation{Institut de
Th\'eorie des Ph\'enom\`enes Physiques, \'Ecole Polytechnique
F\'ed\'erale de Lausanne,  \\ CH-1015 Lausanne,
SWITZERLAND}
\author{G. Lazarides}
\email{lazaride@eng.auth.gr} \affiliation{Physics Division, School
of Technology, Aristotle University of Thessaloniki, Thessaloniki
54124, GREECE}
\author{C. Pallis}
\email{kpallis@gen.auth.gr} 
\affiliation{Department de F\'isica Te\`orica and IFIC, Universitat 
de Val\`encia-CSIC, E-46100 Burjassot, SPAIN}

\date{\today}

\begin{abstract}

A simple extension of the minimal left-right symmetric supersymmetric 
grand unified theory model is constructed by adding two pairs of 
superfields. This naturally violates the partial Yukawa unification 
predicted by the minimal model. After including supergravity 
corrections, we find that this extended model naturally supports hilltop 
F-term hybrid inflation along its trivial inflationary path with only 
a very mild tuning of the initial conditions. With a convenient choice 
of signs of the terms in the \Ka, we can reconcile the inflationary 
scale with the supersymmetric grand unified theory scale. All the 
current data on the inflationary observables are readily reproduced. 
Inflation is followed by non-thermal leptogenesis via the 
decay of the right-handed neutrinos emerging from the decay of the 
inflaton and any possible washout of the lepton asymmetry is avoided 
thanks to the violation of partial Yukawa unification. The extra 
superfields also assist us in reducing the reheat temperature so as to 
satisfy the gravitino constraint. The observed baryon asymmetry of the 
universe is naturally reproduced consistently with the neutrino 
oscillation parameters. 

\end{abstract}

 \maketitle

\section{Introduction}\label{sec:intro}

One of the most natural and well-motivated inflation models is, 
certainly,
the \emph{supersymmetric} (SUSY) \emph{F-term hybrid inflation}
(FHI) \cite{hybrid,susyhybrid}. It is realized at (or close to)
the SUSY \emph{grand unified theory} (GUT) scale $M_{\rm GUT}
\simeq 2.86\times10^{16}~{\rm GeV}$ and can be easily linked to
extensions \cite{lectures} of the \emph{minimal supersymmetric
standard model} (MSSM) which provide solutions to a number of
problems of the MSSM. Namely, the $\mu$-problem of MSSM can be 
solved
via a direct coupling of the inflaton to the electroweak Higgs 
doublet superfields \cite{dvali} or via a \emph{Peccei-Quinn} (PQ) 
symmetry \cite{rsym, pqhi}, which also solves \cite{pq} the strong 
CP problem. Also, baryon-number conservation can be an automatic
consequence \cite{dvali} of a R symmetry and the \emph{baryon 
asymmetry of the universe} (BAU) can be generated via non-thermal 
leptogenesis \cite{ntlepto}, which takes place through the 
out-of-equilibrium decay of the decay products of the inflaton.

Trying to embed SUSY FHI into a concrete SUSY GUT model, we face 
the following challenges: {\it (i)} the possible production of 
topological defects \cite{kibble,monopole} during the GUT phase 
transition at the end of FHI, which in the case of magnetic monopoles 
or domain walls is cosmologically disastrous, {\it (ii)} the mismatch 
\cite{hinova} between the inflationary scale and the SUSY GUT scale, 
and {\it (iii)} the possible washout of the generated lepton-number 
asymmetry due to the smallness of the lightest right-handed neutrino
mass dictated by the various types of  {\it Yukawa unification} (YU) 
conditions predicted by some GUT models.

Here we present a model based on the left-right symmetric GUT gauge
group $G_{\rm LR}={\rm SU(3)_c}\times {\rm SU(2)_L}\times {\rm
SU(2)_R}\times {\rm U(1)}_{B-L}$, which aims to surpass the
problems mentioned above. Namely, $G_{\rm LR}$ does not lead to 
production of magnetic monopoles as higher gauge groups, such 
as the Pati-Salam group, do. Moreover, invoking higher order
terms in the K\"ahler potential with a suitable arrangement of
their signs, as done in \cref{rlarge}, we succeed to overcome the
second of the aforementioned difficulties of SUSY FHI. It is 
important to note that the same form of the K\"ahler potential has 
been proposed in order to re\-concile the value of the scalar 
spectral index $n_{\rm s}$ obtained within SUSY FHI with the present 
data \cite{wmap,plin}. 

Finally and probably most importantly, the problem (iii) is overcome 
by conveniently extending the superfield content of the simplest -- 
see e.g. \cref{vlachos} -- GUT model based on $G_{\rm LR}$. Namely, 
we introduce a pair of ${\rm SU(2)_L}\times {\rm SU(2)_R}$ bidoublet 
superfields and a pair of ${\rm SU(2)_{\rm R}}$ triplet superfields, 
which lead to a sizable violation of the neutrino-$\tau$ (and 
top-bottom) YU predicted by the simplest model. As a consequence, 
the lightest right-handed neutrino mass, which depends heavily on 
the lightest neutrino Dirac mass, may become large enough so that 
any washout of the pre-generated lepton asymmetry is elegantly evaded. 
Moreover, the ${\rm SU(2)_{\rm R}}$ triplet superfields enter the 
inflationary sector of the model leading to a variety of possible 
inflationary scenarios -- see Refs.~\cite{nshift,nsmooth,ax2,ax3} -- 
as well as to extra contributions to the radiative corrections on 
the inflationary paths used in these scenarios. Here we choose to 
analyze FHI along the trivial inflationary trajectory of this model. 
We should note that these same triplet superfields assist us to 
reduce the predicted reheat temperature to an acceptable level.
 
Imposing, in addition, a number of theoretical and observational 
constraints originating from the data on the inflationary 
observables, the boundedness below of the inflationary potential, 
the observed BAU, the gravitino constraint \cite{gravitino,kohri}, 
and the data on the neutrino oscillation parameters, we find a 
wide and natural allowed space of parameters. The resulting FHI 
inflationary scenario is of the hilltop type \cite{lofti} requiring 
a mild tuning of the initial conditions \cite{gpp} to yield 
acceptable values of the scalar spectral index and a rather large 
value of the gravitino mass to fulfill the gravitino constraint.

In Sec.~\ref{theory}, we present the basic ingredients of our
model, while, in Sec.~\ref{fhi}, we describe the inflationary
scenario. We then discuss the inflationary requirements and their
implications for the model parameters in Sec.~\ref{const}. Our 
next step is to outline the mechanism of non-thermal leptogenesis 
in Sec.~\ref{lepto} and update the constraints on the model 
parameters taking into account the post-inflationary requirements 
in Sec.~\ref{cont2}. We summarize our conclusions in 
Sec.~\ref{concl}. Finally, in Appendix A, we present a numerical
analysis of the reheating process in our model.

\section{The SUSY Left-Right Symmetric Model} \label{theory}

We will outline the salient features of our model in 
Sec.~\ref{fields} and analyze the various parts of its 
superpotential in Sec.~\ref{superpot}. Finally, in \Sref{th3}, 
we will derive a set of Yukawa quasi-unification conditions 
which play a key role in our model.

\subsection{Superfield Content and Symmetries} \label{fields}

As already mentioned, we adopt the left-right symmetric gauge 
group $G_{\rm LR}={\rm SU(3)_c}\times {\rm SU(2)_L}\times 
{\rm SU(2)_R}\times {\rm U(1)}_{B-L}$. This gauge group is 
broken down to the \emph{standard model} (SM) gauge group 
$G_{\rm SM}$ 
at a scale close to the SUSY GUT scale $M_{\rm GUT}$ through the 
\emph{vacuum expectation values} (VEVs) acquired by a conjugate 
pair of ${\rm SU}(2)_{\rm R}$ doublet left-handed Higgs 
superfields $\Phi$ and $\bar\Phi$ with $B-L=1,~-1$ respectively. 
In this model, no magnetic monopoles \cite{monopole} or cosmic 
strings \cite{kibble} are produced \cite{trotta} at the end of 
inflation 
and, therefore, we are not obliged to modify \cite{smooth,shifted} 
the standard realization of SUSY FHI to avoid monopole production, 
or impose extra restrictions on the parameters -- as e.g. in 
Ref.~\cite{mairi}.

The representations and transformations under $G_{\rm LR}$ of the
various matter and Higgs superfields of the model are presented in
Table~\ref{tab:fields} ($U_{\rm c}\in {\rm SU(3)_c}$, 
$U_{\rm L}\in {\rm SU(2)_L}$, $U_{\rm R}\in {\rm SU(2)_R}$ and 
$\tr$, $\dagger$, and $\ast$ stand for the transpose, the 
\emph{Hermitian conjugate} (H.c.), and the complex conjugate 
of a matrix respectively). The model also possesses three global 
${\rm U(1)}$ symmetries, namely a R symmetry, a PQ symmetry, and 
the baryon-number ($B$) symmetry. The corresponding charges 
are shown in Table~\ref{tab:fields} too. Note, in passing, that
such continuous global symmetries can effectively arise 
\cite{laz1} from the discrete symmetries emerging in many 
compactified string theories (see e.g. Ref.~\cite{laz2}).  

\begin{table}[!t]
\caption{The superfield content of the model.}
\begin{tabular}{c@{\hspace{0.4cm}}c@{\hspace{0.4cm}}
c@{\hspace{0.4cm}} c@{\hspace{0.4cm}} c@{\hspace{0.4cm}}c}
\toprule
{Super-}&{Represen-}&{Transfor-}&\multicolumn{3}{c}{Global}
\\{fields}&{tations}&{mations}
&\multicolumn{3}{c}{Symmetries}
\\ {}&{under $G_{\rm LR}$}&{under
$G_{\rm LR}$}&{$R$} &{$ PQ$} &{$B$}
\\\colrule
\multicolumn{6}{c}{Matter Fields}\\\colrule
{$l_i$} &{$({\bf 1, 2, 1}, -1)$}& $l_i U_{\rm L}^\dagger$&$1$ & $-1$
&$0$
 \\
{$l^c_i$} & {$({\bf 1, 1, 2}, 1)$} &$U_{\rm R}^\ast
l^c_i$&$1$&{$0$}&{$0$}
\\
{$q_i$} &{$({\bf 3, 2, 1}, 1/3)$}&$q_i U_{\rm L}^\dagger\ U_{\rm
c}^\tr$& $1$ & $-1$ &$1/3$
 \\
{$q^c_i$} & {$({\bf \bar 3, 1, 2},-1/3)$} &$U_{\rm c}^\ast\ U_{\rm
R}^\ast q^c_i$&$1$ &{$0$}&$-1/3$
\\ \colrule
\multicolumn{6}{c}{Higgs Fields}
\\\colrule
{$\Phi$} &{$({\bf 1, 1, 2},1)$}&$U_{\rm R}^\ast
\Phi$&{$0$}&{$0$} & {$0$} \\
{$\bar\Phi$}&$({\bf 1, 1, 2},-1)$& $\bar\Phi U^\tr_{\rm R}$&
{$0$}&{$0$}&{$0$} \\
{$S$} & {$({\bf 1, 1, 1},0)$}&$S$&$2$ &$0$ &$0$ \\ \colrule
{$\hh$} & {$({\bf 1, 2, 2},0)$}&$ U_{\rm L}\hh U^\tr_{\rm R}$&$0$ &$1$
&$0$\\\colrule
{$N$} &{$({\bf 1, 1, 1},0)$}&$N$& {$1$}&{$-1$} & {$0$}\\
{$\bar N$}&$({\bf 1, 1, 1},0)$&$\bar N$& {$0$}&{$1$}&{$0$}
\\\colrule
\multicolumn{6}{c}{Extra Higgs Fields}
\\\colrule
$\hh^{\prime}$&{$({\bf 1, 2, 2},0)$}&$ U_{\rm L}\hh^{\prime}U^\tr_{\rm R}$&
$0$ & $1$ &$0$
\\
$\bhh^{\prime}$&{$({\bf 1, 2, 2},0)$}&$U_{\rm L}\bhh'U^\tr_{\rm R}$& $2$ &
$-1$ &$0$
\\\colrule
$T$&$({\bf 1, 1, 3},0)$ &$U_{\rm R} T U_{\rm R}^\dagger $& $0$ & $0$ &$0$
\\
$\bar T$&{$({\bf 1, 1, 3},0)$}  &$U_{\rm R}\bar T U_{\rm R}^\dagger $& $2$ &
$0$ &$0$
\\\botrule
\end{tabular}\label{tab:fields}
\end{table}

The lepton and quark superfields are $l_i$, $l^c_i$ and $q_i$, 
$q^c_i$ ($i=1,2,3$) respectively -- we follow here the same
representation of the superfields under ${\rm SU(2)_L}\times 
{\rm SU(2)_R}$ as in Ref.~\cite{klpnova}. In the simplest 
version of the model without the extra Higgs superfields in 
Table~\ref{tab:fields}, the electroweak Higgs doublets $H_1$ and 
$H_2$ which couple to the down- and up-type quarks respectively 
belong to the bidoublet superfield $\hh$. So, as one can easily 
see, all the requirements \cite{pana} for partial YU, i.e. the 
`asymptotic' (at $M_{\rm GUT}$) equality of the Yukawa coupling 
constants of the $t$- and the $b$-quark as well as of the 
$\tau$-neutrino $\nu_\tau$ and the $\tau$-lepton $\tau$, are 
fulfilled. As already indicated, the breaking of $G_{\rm LR}$ 
down to $G_{\rm SM}$ is achieved by the superheavy VEVs 
($\sim M_{\rm GUT}$) of the conjugate pair of Higgs superfields 
$\phh$, $\bar{\phh}$ along their right-handed neutrino type 
components ($\nu^c_\phh$, $\bar{\nu}^c_\phh$). 
The model also contains a gauge singlet $S$, which triggers the 
breaking of $G_{\rm LR}$ and a pair of gauge singlets $N$, 
$\bar{N}$ for solving \cite{rsym} the $\mu$ problem of the MSSM 
via the PQ symmetry.

The partial YU between the $b$- and the $t$-quark implied by 
the simplest left-right symmetric model is not compatible
\cite{gYqu,klpnova,Antuch} with the constrained MSSM (CMSSM), 
which is based on universal boundary conditions for the soft 
SUSY breaking parameters. Actually, a sizable violation of 
partial YU is required within the context of the CMSSM, which 
we adopt here. In order to achieve this violation, we extend 
the model by including four extra Higgs superfields $\hh'$, 
$\bhh'$, $T$, and $\bar T$, where the barred superfields are 
included in order to give superheavy masses to the unbarred 
superfields. These extra Higgs superfields together with 
their transformation properties and charges are also included 
in Table~\ref{tab:fields}. The superfield $\hh'$ belongs to 
the ({\bf 1,2,2},0) representation of $G_{\rm LR}$ and, 
therefore, can couple to the fermions. The triplet $T$ 
acquires a superheavy VEV of order $M_{\rm GUT}$ after the 
breaking of $G_{\rm LR}$ to $G_{\rm SM}$. Its couplings with 
$\bhh^{\prime}$, $\hh^{\prime}$, and $\hh$ then naturally generate a 
${\rm SU(2)_R}$-violating mixing of the ${\rm SU(2)_L}$ doublets in 
$\hh$ and $\hh^{\prime}$ leading, thereby, to a sizable violation of 
partial YU.

\subsection{Superpotential Terms} \label{superpot}

The superpotential $W$ of our model can be split into three parts:
\beq W=W_{\rm H}+W_{\rm m}+W_{\rm Y}+W_{\rm NR}, 
\label{Wtot}\eeq
which are analyzed in the following.

\paragraph{} $W_{\rm H}$ is the part of the superpotential which is 
relevant for the breaking of $\Ggut$ to $\Gsm$ and is given by
\beq W_{\rm H}=\kappa S\lf \bph\phh-M^2\rg-\kp_TST^2+
M_T\bar{T}T+\ld \bar T\bph\phh\label{WH}, \eeq
where the mass parameters $M$ and $M_T$ are of order 
$M_{\rm GUT}$, and $\kappa$, $\kp_T$, and $\lambda$ are 
dimensionless parameters. Note that $M$, $M_T$, $\kappa$, and 
$\ld$ can be made real and po\-sitive by field redefinitions.
The third dimensionless parameter $\kp_T$, however, remains in
general complex. For definiteness, we choose this parameter to be
real too, but of any sign. The parameters are normalized so that 
they correspond to the couplings between the SM singlet 
components of the superfields.

\par
The scalar potential obtained from $W_{\rm H}$ is given by
\begin{eqnarray}
V_{\rm H}&=&\left\vert\kappa(\bph\phh-M^2)-\kp_T T^2 \right\vert^2
+\left\vert 2\kp_T ST-M_T\bar{T} \right\vert^2 \nonumber \\
%
%
&&+\left\vert\kappa S+\lambda\bar{T} \right\vert^2\left(\vert
\phh\vert^2+\vert\bar{\phh}\vert^2\right)
+\left\vert M_T T+\lambda \bph\phh \right\vert^2\nonumber \\
&&+\ {\rm D-terms}, \label{potential}
\end{eqnarray}
where the complex scalar fields which belong to the SM singlet
components of the superfields are denoted by the same symbols as
the corresponding superfields.  Vanishing of the D-terms yields
$\bar{\phh}\,^{*}=e^{i\vartheta}\phh$ ($\phh$, $\bph$ lie in the
$\nu^c_\phh$, $\bar{\nu}^c_\phh$ direction),  where $\vartheta$ 
is an arbitrary phase. Performing appropriate R and gauge 
transformations, we bring $\phh$  and $S$ to the positive real 
axis, while $\bph$ stays in general complex with a phase factor
$e^{-i\vartheta}$.

We define a combination of the five real parameters of the model
\beq \label{xdef}
\xi={\frac{\kp_T\lambda^2}{\kappa}}{\frac{M^2}{M_T^2}}. 
\eeq 
From the potential in \Eref{potential}, one can then show that,
under the assumption that $\xi<1/4$, the nearest to the trivial 
flat direction (see below) SUSY vacuum, where the system is most 
likely to end up after the end of inflation, corresponds to 
$\vartheta=0$ (for both signs of $\xi$) and lies at
\beqs\beq \label{vev1}
\vev{S}=\vev{\bar{T}}=0,~~\vev{\bph\phh}=v^2_\phh,~~
\vev{T}=v_T\lf1,1,\frac{\sigma_3}{\sqrt{2}}\rg, 
\eeq 
where
\beq
\lf{\frac{v_\phh}{M}}
\rg^2=\frac{1}{2\xi}\left(1 - \sqrt{1-4\xi}\right),~~
v_T=-{\ld \frac{v_\phh^2}{M_T}}, \label{vev2} \eeq\eeqs
and $\sigma_{3}={\sf diag}\lf1,-1\rg$.

\paragraph{} $W_{\rm m}$ is the part of the superpotential which is 
responsible for the mixing of the doublets in $\hh$ and $\hh'$ and 
can be written symbolically as
\beq W_{\rm m}=m\bhh'\hh+ m'\bhh'\hh^{\prime}+\ldt T\bhh'\hh
+\ldt' T\bhh'\hh',\label{Wm}\eeq
where the mass parameters $m$ and $m'$ are of order $M_{\rm GUT}$
(made real and positive by field rephasing) and $\ldt$, $\lds$ are
dimensionless complex coupling constants. Note that the two last
terms in the \emph{right hand side} (RHS) of \Eref{Wm} overshadow 
the corresponding ones from the non-renormalizable 
${\rm SU(2)_R}$-triplet couplings originating from the symbolic 
couplings
$\bar{\Phi}\Phi\bhh^{\prime}\hh$ and $\bar{\Phi}\Phi\bhh^{\prime}
\hh^{\prime}$ -- see Ref.~\cite{qcdm}. 
\paragraph{} $W_{\rm Y}$ contains the Yukawa interactions of 
the fermions and is given by
\beq  W_{\rm Y}=q_i\left(y_{ijQ}\hh+y'_{ijQ}\hh'\right)q_j^c+
l_i\left(y_{ijL}\hh+y'_{ijL}\hh'\right)l_j^c, \label{Wy}\eeq
where $y_{ijQ}$ and $y_{ijL}$ are, respectively, the Yukawa 
coupling constants of the quarks and lepton with the Higgs 
superfield $\hh$, while $y'_{ijQ}$ and $y'_{ijL}$ are their 
Yukawa coupling constants with $\hh'$.
 
Defining properly \cite{qcdm,klpnova} the symbolic couplings in 
the RHS of \Eref{Wm}, we obtain the mass terms
\bea\nonumber W_{\rm m}&=&\left(m^\prime-\frac{\ldt' v_T}
{\sqrt{2}}\right)\left(\hh^{\prime \tr}_1+\alpha_1\hh^\tr_1
\right)\openep\bhh^{\prime}_2
\\
&&+\left(m^\prime+\frac{\ldt' v_T}{\sqrt{2}}
\right)\bar\hh^{\prime\tr}_1\openep\left(\hh^{\prime}_2+
\alpha_2\hh_2\right)
+\cdots,~~~
\label{superheavy}\eea
where $\openep$ is the $2\times 2$ antisymmetric matrix with
$\openep_{12}=1$, the ellipsis includes color non-singlet
components of the superfields, and the complex dimensionless
parameters $\alpha_{1}$ and $\alpha_{2}$ are given by
\beqs\bea\label{alphas1}
\alpha_{1}&=&\frac{m-\ldt v_T/\sqrt{2}}{m^\prime-\ldt' v_T/
\sqrt{2}}, \\ 
\alpha_{2}&=&\frac{m+\ldt v_T/\sqrt{2}}{m^\prime+\ldt' v_T/
\sqrt{2}}\cdot \label{alphas2} \eea\eeqs

\paragraph{} $W_{\rm NR}$ is the part of $W$ which
contains its non-renormalizable terms:
\bea \nonumber W_{\rm NR}&=&\ld_{ij}\frac{\bph\bph
l_i^{c}l_j^{c}}{\Ms}+\ld_{N}\frac{N^2\bar N^2}{\Ms}\\ 
&&+\frac{N^2}{2\Ms}\lf\lm\hh^2+\lm'\hh\hh'+\lm''\hh^{\prime2}
\rg+\cdots,~~~~\label{Wnr}\eea
where $M_{\rm S}\simeq 5\cdot 10^{17}~{\rm GeV}$ is an effective
scale comparable to the string scale. Here we have displayed 
expli\-citly only the terms which are relevant for our analysis.
The first term in the RHS of this equation is responsible for
generating intermediate scale Majorana masses for the 
right-handed neutrinos after the breaking of $G_{\rm LR}$. 
These masses together with the Dirac neutrino masses in 
Eq.~(\ref{mDirac}) lead to the light neutrino masses via the 
seesaw mechanism. The same term is important for the decay of
the inflaton system after the end of inflation to right-handed 
neutrinos and sneutrinos, whose subsequent decay can lead to 
non-thermal leptogenesis. The fact that this term is suppressed
by $M_{\rm S}$ guarantees a sufficiently low reheat temperature
which is useful for a successful leptogenesis -- see \Sref{lepto}. 
Finally, the second and third term provide the $\mu$ term of MSSM
along the lines of Ref.~\cite{rsym}.

\subsection{Yukawa Quasi-Unification Conditions} \label{th3}

It is obvious from Eq.~(\ref{superheavy}) that we obtain two
pairs of superheavy doublets:
\beqs\beq\label{shs1}\bhh^{\prime}_1,~H^{\prime}_2~~\mbox{and}
~~H^{\prime}_1,~\bhh^{\prime}_2, \eeq where \beq\label{shs11}
H^{\prime}_{r}=\frac{\hh^{\prime}_{r}+\alpha_{r}\hh_{r}}
{\sqrt{1+|\alpha_{r}|^2}},~r=1,2 \eeq\eeqs
(no summation over the repeated index $r$ is implied). The
electroweak doublets $H_r$, which remain massless at the GUT
scale, are orthogonal to the $H^{\prime}_{r}$ directions:
\beq H_r=\frac{-\alpha_r^*\hh^{\prime}_r+\hh_r}
{\sqrt{1+|\alpha_r|^2}} \cdot\label{elws}\eeq
Solving Eqs.~(\ref{shs11}) and (\ref{elws}) with respect to 
$\hh_r$ and $\hh^\prime_r$, we obtain
\beq
\hh_r=\frac{H_r+\alpha^*_rH^{\prime}_r}{\sqrt{1+|\alpha_r|^2}}
~~\mbox{and}~~\hh^\prime_r=\frac{-\alpha_rH_r+H^{\prime}_r}
{\sqrt{1+|\alpha_r|^2}}\cdot~~~\eeq
The superheavy doublets $H^{\prime}_r$ must have zero VEVs,
which gives
\beq\label{hvev}
\vev{\hh_r}=\frac{\vev{H_r}}{\sqrt{1+|\alpha_r|^2}}
~~\mbox{and}~~\vev{\hh^\prime_r}=\frac{-\alpha_r\vev{H_r}}
{\sqrt{1+|\alpha_r|^2}}\cdot~~~\eeq

From Eqs.~(\ref{Wy}) and (\ref{hvev}), we can readily derive the 
mass matrices of the up- and down-type quarks ($m_{ijU}$ and 
$m_{ijD}$ respectively), as well as the Dirac mass matrix 
$m^{\rm D}_{ij\nu}$ of the neutrinos and the mass matrix $m_{ijE}$ 
of the charged leptons:
\beqs\bea && \label{mU}
m_{ijU}=\frac{y_{ijQ}-\alpha_2y'_{ijQ}}
{(1+|\alpha_2|^2)^{\frac{1}{2}}}v_2\equiv \hat{y}_{ijU}v_2,
\\ && \label{mD}
m_{ijD}=\frac{y_{ijQ}-\alpha_1y'_{ijQ}}
{(1+|\alpha_1|^2)^{\frac{1}{2}}}v_1\equiv \hat{y}_{ijD}v_1, 
\\ && \label{mDirac}
m^{\rm D}_{ij\nu}=\frac{y_{ijL}-\alpha_2y'_{ijL}}
{(1+|\alpha_2|^2)^{\frac{1}{2}}}v_2\equiv \hat{y}^{\rm D}_{ij\nu}v_2, 
\\ && \label{mE}
m_{ijE}=\frac{y_{ijL}-\alpha_1y'_{ijL}}
{(1+|\alpha_1|^2)^{\frac{1}{2}}}v_1\equiv \hat{y}_{ijE}v_1,
\eea \eeqs
where $v_r=\vev{H_r}$, $\hat{y}_{ijU}$ and $\hat{y}^{\rm D}_{ij\nu}$ 
are, respectively, the effective Yukawa coupling constants of 
the up-type quarks and the neutrinos with $H_2$, and 
$\hat{y}_{ijD}$ and $\hat{y}_{ijE}$ are, respectively, the 
effective Yukawa coupling constants of the down-type quarks and 
the charged leptons to $H_1$. 

In the absence of the superfields $T$ and $\bar T$ which generate 
the ${\rm SU(2)_R}$-violating mixing of the doublets in $\hh$ and 
$\hh'$, Eqs.~(\ref{alphas1}) and (\ref{alphas2}) imply that
$\alpha_1=\alpha_2$. This means that
\beq
\hat{y}_{ijU}=\hat{y}_{ijD}~~\mbox{and}~~\hat{y}^{\rm D}_{ij\nu}=
\hat{y}_{ijE},
\eeq 
i.e. exact asymptotic YU between the up- and down-type quarks 
as well as between the neutrinos and the charged leptons not 
only for the third but for all three families of fermions. In 
particular, there is no mixing in the quark sector. So the 
presence of the $T$ and $\bar T$ superfields is absolutely 
vital for the phenomenological viability of the model. 

Our present analysis is very similar to the analysis in 
Refs.~\cite{qcdm,gYqu,muneg,klpnova,yqu,pekino}, where a set 
of generalized or monoparametric asymptotic Yukawa 
quasi-unification conditions have been obtained. There are, 
however, two important differences. In 
these references, only the third generation of fermions has
been considered and the gauge group was larger than the 
left-right symmetric gauge group $G_{\rm LR}$ used here, 
yielding a relation between the quark and lepton Yukawa 
coupling constants too and allowing the desired mixing of 
the ${\rm SU(2)_L}$ Higgs doublets even with just a pair of 
${\rm SU(2)_R}$ Higgs singlets. In this paper, the quark and 
lepton sectors are completely independent as one can see from 
Eqs.~(\ref{mU}), (\ref{mD}), (\ref{mDirac}), and (\ref{mE}). 
We will not consider further the quark sector here. We will 
rather concentrate on the lepton sector since this sector is 
important for the scenario of non-thermal leptogenesis, which 
is discussed in \Sref{lepto}.   

\section{The Inflationary Scenario}\label{fhi}

In \Sref{fhi1}, we describe the inflationary trajectory and, in
Secs.~\ref{fhi2} and \ref{fhi3}, we present the radiative and
\emph{supergravity} (SUGRA) corrections incorporated in the 
inflationary 
potential. Finally, in \Sref{fhi3}, we extract the inflationary 
observables.

\subsection{The Inflationary Trajectory}\label{fhi1}

The superpotential terms which are relevant for inflation 
constitute $W_H$ in \Eref{WH}. From the derived $F$-term 
scalar potential in \Eref{potential}, we can deduce that 
the model under discussion possesses the following 
classically flat directions:

\begin{itemize}

\item The trivial one, which lies at 
\beqs\beq\Phi=\bar\Phi=T=\bar T=0\label{flatr}\eeq 
with potential energy density 
\beq V^0_{\rm tr}=\kp^2 M^4.\label{Vhio}\eeq 
This is a valley of local minima in the $\Phi$, $\bar\Phi$ 
directions for
\beq |S|>S_{\rm c}\equiv\;M \label{Scr}\eeq
but, for $|S|<S_c$, 
is destabilized in the $(\Phi+\bar\Phi^*)/\sqrt{2}$ 
direction. Let us note, in passing, that, under some 
circumstances, this trajectory, for $|S|<S_c$, gives its 
place to a classically non-flat valley of minima on 
which new smooth FHI can take place along the lines of 
Ref.~\cite{nsmooth}. The $4\times 4$ mass-squared matrix 
$M^2_{T\bar T}$ of the scalar fields $T$, $\bar T$, $T^*$, 
and $\bar T^*$ has determinant and trace
\bea & &{\sf Det}\lf M^2_{T\bar T}\rg =\nonumber\\
& & \mt^4 \lf \mt^2 +2 \kp \kp_T M^2\rg\, 
\lf \mt^2 - 2 \kp \kp_T M^2\rg\label{dettr}\eea 
and
\beq {\sf Tr}\lf M^2_{T\bar T}\rg= 4\lf \mt^2+ 
2 \kp_T^2 S^2\rg\,\label{trtr}\eeq
respectively. It can be easily shown that the mass-squared 
matrix $M^2_{T\bar T}$ of the scalar $T$, $\bar{T}$ system 
has four positive eigenvalues for 
\beq |\kp_T| < \frac{\mt^2}{2 \kp M^2}~~\Rightarrow~~|\xi|
<\frac{\ld^2}{2\kp^2}\label{kteq1}\eeq \eeqs
and, thus, the trivial flat direction is an honest candidate
inflationary trajectory since it is stable in the $T$, 
$\bar T$ scalar field directions for all the values of $S$. 
On the contrary, 
violation of the bound in \Eref{kteq1} implies that at least 
one of the eigenvalues of the mass-squared matrix 
$M^2_{T\bar T}$ is negative and, thus, this direction is a 
path of saddle points for all the values of the field $S$. 
In this case, another inflationary path comes into existence, 
namely the semi-shifted one. 

\item The semi-shifted path found at $\phh=\bph=0$ and
\beqs\beq
T=\pm\sqrt{-\frac{\kp}{\kp_T}M^2-\frac{\mt^2}{2\kp_T^2}},~~
\bar T=\frac{2\kp_T}{\mt}ST \label{semishift} \eeq
with \beq|\kp_T| > \mt^2/(2 \kp M^2)\label{kteq2}\eeq
and potential energy density
\beq V^0_{\rm ssh}=-\frac{\mt^4 + 4 \kp \kp_T \mt^2 M^2}{4
\kp_T^2}\cdot\eeq\eeqs
On this path, the left-right symmetric gauge group 
$G_{\rm LR}$ is broken to $G_{\rm SM}\times 
{\rm U}(1)_{B-L}$ and a semi-shifted FHI can occur as 
shown in Ref.~\cite{ax3}.

\item The shifted path, which appears at
\beqs\bea
&& \bph\phh=\kp\frac{(\kp^2 + 2 \ld^2) \mt^2 + 4 \kp
\kp_T \ld^2 M^2}{4 \kp_T \ld^2 (\kp^2 + \ld^2)},\label{nsh1}\\
&& T=-\frac{\kp\mt}{2\ld\kp_T},\quad \bar T=-\frac{\kp S}{\ld} 
\label{nsh2}
\eea
with potential energy density 
\bea V^0_{\rm nsh}=\frac{\kp^2 (\kp
\mt^2 - 4 \kp_T \ld^2 M^2)^2}{16 \kp_T^2 \ld^2 (\kp^2 +
\ld^2)}\label{Vnsh}.\eea \eeqs
This trajectory is analogous to the one used for the new shifted
FHI of Ref.~\cite{nshift}. Along this direction, $G_{\rm LR}$ is
broken to $G_{\rm SM}$.

\end{itemize}

\renewcommand{\arraystretch}{1.4}

\begin{table*}[!t]
\caption{The mass spectrum of the model along the inflationary
trajectory in \Eref{flatr}.}
\begin{tabular}{c|@{\hspace{0.5cm}}c@{\hspace{0.5cm}}|@
{\hspace{0.5cm}} c @{\hspace{0.5cm}}|@{\hspace{0.5cm}}c@
{\hspace{0.5cm}}|@{\hspace{0.5cm}} c}\toprule
{Superfields} &{Real} & {Masses}&{Weyl} & {Masses}\\
{of Origin} & {Scalars}&{} & {Spinors}&{}\\\colrule
$\Phi$, $\bar\Phi$ & $2\times 4$  & $m_{{\bf 2}\pm}=\kp(\lfa
S\rga^2\pm M^2)^{1/2}$ &$2\times 2$&$M_{{\bf 2}\pm}=\pm\kp\lfa
S\rga$\\\colrule
$\bar T$, $T$ & $3\times 2$ & $m_{{\bf 3}\pm}=\lf\mps^2
\pm\sqrt{\Dp}\rg^{1/2} $& $3\times 2$ & $M_{{\bf 3}\pm}=
\pm\kpt\lfa S\rga+$\\
 & $3\times2$ & $\bar m_{{\bf 3}\pm}=\lf\mng^2\pm
 \sqrt{\Dn}\rg^{1/2}  $&&$\sqrt{\mt^2+\kpt^2\lfa S\rga^2}$
\\\botrule
\end{tabular}
\label{tab1}
\end{table*}

In our subsequent discussion, we will impose the condition in 
\Eref{kteq1} and concentrate on the first case above, where 
the semi-shifted flat direction in Eq.~(\ref{semishift}) does 
not exist. Writing the potential energy density $V^0_{\rm nsh}$ 
in Eq.~(\ref{Vnsh}) in the form
\beq V^0_{\rm nsh}=\frac{\ld^2}{\kp^2+\ld^2}\lf\frac{1}{4\xi}-
1\rg^{2}V^0_{\rm tr}, \label{Vnsh1}\eeq
we can show that
\beq
V^0_{\rm nsh}>V^0_{\rm tr}\label{Vneq1}\eeq 
for 
\beq
\frac{1}{4\lf1-{\sqrt{\kp^2+\ld^2}/\ld}\rg}<
\xi<\frac{1}{4\lf1+{\sqrt{\kp^2+\ld^2}/\ld}\rg}. \label{xineq}
\eeq
Under these circumstances, it is more likely that the system 
will eventually settle down on the trivial rather than the 
new shifted flat direction and will undergo FHI of the 
standard type along the trivial path. In the opposite case, 
where $V^0_{\rm nsh}<V^0_{\rm tr}$, we better ensure that 
the critical value $S_{\rm nc}$ of $S$ on the new shifted 
path in \eqs{nsh1}{nsh2} is larger than the critical $S$ on 
the trivial path given in \Eref{Scr}. In this case, the 
system, after the end of inflation along the trivial path 
in \Eref{flatr}, is expected to fall directly into the SUSY 
vacuum
without being trapped in the shifted path, where it 
could undergo a second stage of inflation. Taking into 
account the findings of \cref{nshift}, we see that the last 
prerequisite is achieved if 

\begin{widetext}
\beq \frac{S_{\rm nc}}{M}\equiv\lfa \frac{\kp \ld^2 \lf{1}/{4
\xi}-1\rg \lf2 \kp^2 \lf1 + (\kp + 2 \kpt)/{4 \xi \kp}\rg + {(\kp
+ \kpt) \ld^2}/{\xi \kp}\rg}{2\kpt (\kp^2 + \ld^2) \lf 2 \lf1 + {1}/{4
\xi}\rg \kp^2 + {\ld^2/\xi}\rg}\rga^{1/2}>1.\label{Sncr}\eeq
\end{widetext}

\subsection{Radiative Corrections}\label{fhi2}

The constant tree-level potential energy density $\Vhio\equiv
V_{\rm tr}^0$, which drives inflation along the trivial trajectory, 
causes SUSY breaking leading \cite{susyhybrid} to the generation of 
one-loop radiative corrections, which provide a logarithmic slope 
along the inflationary path. To calculate these corrections, we 
construct the mass spectrum of the theory on the inflationary path 
in Eq.~(\ref{flatr}). Our results are summarized in Table~\ref{tab1}, 
where we have defined
\beqs \beq m_{\pm}^2\equiv \mt^2\pm \kp \kpt M^2 + 2\kpt^2 \lfa
S\rga^2\label{mpm}\eeq and \beq D_{\pm}\equiv  \kpt^2 \lf4\mt^2
\lfa S\rga^2 + \lf \kp M^2 \pm 2\kpt \lfa
S\rga^2\rg^2\rg.\label{Dpm} \eeq\eeqs
As we anticipated in the first item of \Sref{fhi1}, we see, from 
Table~\ref{tab1}, that
the mass-squared matrix of the scalar components of the $\Phi$ 
and $\bar\Phi$ superfields develops a negative eigenvalue as $|S|$ 
crosses below its critical value $S_{\rm c}$, whereas the system 
of the scalar components of the $T$ and $\bar T$ is completely 
stable for all values of $S$ provided that the condition in 
\Eref{kteq1} is satisfied.

Inserting the spectrum shown in Table~\ref{tab1} in the well-known
Coleman-Weinberg formula \cite{cw}, we find that the one-loop
radiative correction to $\Vhio$ is
\begin{equation} \label{Vcor} V_{\rm HIc}=V_{\bar\Phi\Phi}+
V_{\bar TT},\eeq
where
\beqs\beq \label{Vpp} V_{\bar\Phi\Phi} =
\frac{2}{64\pi^2}\sum_{I=+,-}\left(2m_{{\bf2}I}^4
\ln\frac{m_{{\bf2}I}^2}{\Lambda^2}-2M_{{\bf2}I}^4
\ln\frac{M_{{\bf2}I}^2}{\Lambda^2}\right)\eeq
and \bea V_{\bar TT}&=&
\frac{3}{64\pi^2}\sum_{I=+,-}\left(m_{{\bf3}I}^4
\ln\frac{m_{{\bf3}I}^2}{\Lambda^2}+\bar m_{{\bf3}I}^4
\ln\frac{\bar m_{{\bf3}I}^2}{\Lambda^2}\right.
\nonumber\\
&&\left.-2M_{{\bf3}I}^4\ln\frac{M_{{\bf3}I}^2}{\Lambda^2}
\rg\label{Vtt} \eea\eeqs
with $\Lambda$ being a renormalization scale. In the relations
above, we have taken into account that the dimensionality of the
representations to which ${\Phi}$, $\bar\Phi$  and $T$, $\bar T$ 
belong is 2 and 3 respectively -- see Table~\ref{tab:fields}. It 
is important to note that
\beqs\beq\sum_{I=+,-}\lf2 m_{{\bf2}I}^4 -2M_{{\bf2}I}^4\rg=4\kp^4
M^4\eeq and \beq\sum_{I=+,-}\left( m_{{\bf3}I}^4 +\bar
m_{{\bf3}I}^4 -2M_{{\bf3}I}^4\rg=8\kp^2\kpt^2M^4\eeq \eeqs
are $S$-independent, which implies that the slope of the 
inflationary trajectory is
$\Lambda$-independent and the scale $\Lambda$, which remains
undetermined, does not enter the inflationary observables.
Moreover, we can show that, in the limit $x=\sigma^2/2M^2\gg1$,
the potential $V_{\rm HIc}$ in \Eref{Vcor} can be well 
approximated by
\beqs\beq V_{\rm HIc}\simeq\Vhio\lf2\kappa^2f_{\rm
rc}(\kp^2x)+6\kpt^2f_{\rm rc}(4\kpt^2x)\rg, \label{Vhica} \eeq 
where 
\beq
f_{\rm rc}(z)=\frac{1}{16\pi^2}\lf \ln\frac{z M^2}{
\Lambda^2}+\frac{3}{2}\rg\cdot\eeq\eeqs
As can be easily deduced from these formulas, $V_{\rm HIc}$ 
is independent of $\ld$ and the sign of $\kpt$ and, to a 
considerable degree, of $\mt$ too.

\subsection{Supergravity Corrections}\label{fhi3}

The F-term tree-level SUGRA scalar potential $\Vhios$ of our
model on the trivial path is obtained from $W_{\rm H}$ in 
Eq.~(\ref{WH}) and the K\"{a}hler potential $\Khi$ by applying 
the standard formula
\beqs\bea \Vhios=e^{\Khi/\mP^2}\left(K^{\aal\bt}{\rm F}^*_\aal
{\rm F}_\bt -3\frac{\vert W_{\rm HI}\vert^2}{\mP^2}\right)
\label{Vsugra} \eea 
with 
\bea
K_{\al\bbet}=\frac{\partial^2\Khi}{\partial\phi^\al\partial
\phi^{*\bbet}},~~
K^{\aal\bt}K_{\bt\bar \gamma}=\delta^\aal_{\bar \gamma},
\label{Vsugra1}\eea
and \bea {\rm F}_\al=\frac{\partial W_{\rm HI}}{\partial\phi^\al} 
+\frac{\partial K}{\partial\phi^\al}\frac{W_{\rm HI}}{\mP^2}, 
\label{Vsugra2}\eea\eeqs
where $\mP$ is the reduced Planck scale and $\phi^\al$ denotes
the complex scalar fields of the model with $\phi^{*\aal}$ 
being their complex conjugates. The K\"{a}hler potential is a
real function of the complex scalar fields and their complex 
conjugates and must respect all the symmetries of the model 
presented in Table~\ref{tab:fields} (including the R symmetry). 
We consider here a generic form of the K\"{a}hler potential, 
which, however, does not deviate very much from the canonical 
one and can, thus, be expanded as follows: 
\bea\nonumber
K&=&\p^2+|\phh|^2+|\bph|^2+\Tr|T|^2+\Tr|\bar T|^2\\
\nonumber
&&+\frac{1}{4}\ks\frac{\p^4}{\mP^2}+
\frac{1}{6}\kss\frac{\p^6}{\mP^4}
+\frac{1}{8}\ksss\frac{\p^8}{\mP^6} \\
&&+\frac{1}{10}\kst\frac{\p^{10}}{\mP^8}
+\frac{1}{12}\ksv\frac{\p^{12}}{\mP^{10}}+\cdots,
\label{K} \eea
where $\ks$, $\kss$, $\ksss$, $\kst$, and $\ksv$ are real 
positive or negative constants of order unity and the ellipsis 
represents terms of higher order involving only the inflaton 
field $S$ as well as terms of higher order in the waterfall 
fields $\Phi$, $\bar\Phi$, $T$, and $\bar T$ and any order in 
$S$. We neglect the latter terms since, as we will now show, 
they are irrelevant on the trivial inflationary path (the 
minimal terms for the waterfall fields are also irrelevant
during inflation, but we include them in the expansion since
they are necessarily present). 

To prove this statement, observe from Table~\ref{tab:fields} 
that the symmetries of the model do not allow terms in $K$ 
which are linear in the waterfall fields. So the only terms in 
$K$ involving these fields are quadratic or of higher order in 
these fields. From Eq.~(\ref{Vsugra2}), we then see that these 
terms do not contribute to ${\rm F}_\al$ evaluated on the 
trivial path. The only way for terms in $K$ involving waterfall 
fields to contribute to the potential on the trivial path is 
then via $K^{\aal\bt}$. However, even this does not happen for 
the following reason. It is clear that $K_{\al\bbet}$ vanishes 
on the trivial inflationary trajectory if just one of its 
indices corresponds 
to a waterfall field, which implies the same property for 
$K^{\aal\bt}$ too. Consequently, the terms in $K$ involving 
waterfall fields could influence the inflationary potential 
only via $K^{\aal\bt}$ with both its indices corresponding to 
waterfall fields. However, these are multiplied by ${\rm F}_\al$ 
with $\al$ corresponding to waterfall fields, which are zero on 
the trivial trajectory as one can see from Eqs.~(\ref{WH}) and 
(\ref{Vsugra2}). 

Using Eqs.~(\ref{WH}), (\ref{Vsugra}), and (\ref{K}), the SUGRA 
scalar potential $\Vhios$ on the trivial trajectory can be 
expanded as follows:
\beq\label{Vol} \Vhios\simeq V_{\rm HI0}
\left(1+\sum_{\nu=1}^5(-1)^{\nu}c_{2\nu K}\lf\frac{\sgm}
{\sqrt{2}\mP}\rg^{2\nu}\right),\eeq
where $\sigma=\sqrt{2}S$ is the real inflaton field which is 
canonically normalized (neglecting terms of order $\p^2$ or 
higher which multiply the kinetic term of $S$) with $S$ being 
rotated on the real axis by an appropriate R transformation. 
Here
\beqs\bea\label{c2k} c_{2K}&=&\ks,\\
\label{c4k}
c_{4K}&=&\frac{1}{2} - \frac{7 \ks}{4} + \ks^2 - \frac{3\kss}{2},\\
c_{6K}&=&-\frac{2}{3} + \frac{3 \ks}{2} - \frac{7 \ks^2}{4} +
\ks^3 + \frac{10\kss}{3}\nonumber \\&& - 3 \ks\kss + 2\ksss,\label{c6k}\\
c_{8K}&=& \frac{3}{8} - \frac{5 \kst}{2} - \frac{13 \ks}{24} + \frac{41
\ks^2}{32} - \frac{7 \ks^3}{4} + \ks^4 \nonumber
\\&&- \frac{13\kss}{4} + \frac{143 \ks\kss}{24} - \frac{9 \ks^2\kss}{2} +
\frac{9\kss^2}{4}\nonumber \\&& - \frac{39\ksss}{8} + 4 \ks \ksss,
\label{c8k}\\c_{10K}&=&-\frac{2}{15} + \frac{32 \kst}{5} + 3 \ksv +
\frac{\ks}{24} - 5 \kst \ks  \nonumber\\&& - \frac{13 \ks^2}{24} + \frac{41
\ks^3}{32} - \frac{7 \ks^4}{4} + \ks^5 + \frac{5\kss}{3}
\nonumber\\&& - \frac{29 \ks\kss}{6} + \frac{103 \ks^2\kss}{12} - 6
\ks^3\kss - 5\kss^2 \nonumber \\ && + \frac{27 \ks \kss^2}{4} +
5\ksss - \frac{67 \ks\ksss}{8} \nonumber \\&& + 6 \ks^2\ksss- 6\kss
\ksss. \label{c10k}\eea\eeqs
In the sum which appears in the RHS of Eq.~(\ref{Vol}), we have kept 
only the first five terms, i.e. the terms up to the tenth order in 
$\sigma$, which is consistent with the expansion of the K\"{a}hler 
potential $K$ in Eq.~(\ref{K}) up to the twelfth order in $|S|$. Note 
that, although the inflationary observables have a non-negligible 
dependence only on the two or three lower terms in the sum in the 
RHS of Eq.~(\ref{Vol}), we included some of the higher terms too 
since these terms control the asymptotic behavior of the potential 
and are, thus, needed in order to guarantee that the potential 
is bounded below at large values of $|S|$ -- see Sec.~\ref{const}. 

The overall inflationary potential $V_{\rm HI}$ on the trivial path is 
found by adding the SUGRA inflationary potential $V^{\rm SG}_{\rm HI0}$
in Eq.~(\ref{Vol}) and the one-loop radiative correction $V_{\rm HIc}$ 
in Eq.~(\ref{Vcor}): 
\beq
V_{\rm HI}=V^{\rm SG}_{\rm HI0}+V_{\rm HIc}. \label{Vtot}
\eeq 

\section{Constraining the Model Parameters}\label{const}

We will now describe, in \Sref{cont1}, the inflationary constraints 
which we will impose on the resulting cosmological scenario, and 
delineate, in Sec.~\ref{num}, the parameter space of our model which 
is allowed by these constraints.

\subsection{Inflationary Requirements}\label{cont1}

We assume that (i) the observed curvature perturbation is 
solely due to the inflaton field $\sigma$, (ii) $\xi<1/4$ and 
the restrictions in \eqs{kteq1}{Vneq1} or (\ref{Sncr}) are 
fulfilled, and (iii) the FHI is followed by damped coherent 
oscillations about the SUSY vacuum until reheating after 
which radiation dominates leading eventually to matter dominance. 
Under these hypotheses, the parameters of our model can be 
further restricted by imposing the following requirements:

\setcounter{paragraph}{0}

\paragraph{}  The number of e-foldings $\Nhi$ that the pivot 
scale $k_*=0.05/{\rm Mpc}$ undergoes during FHI has to lead 
to a solution of the horizon and flatness problems of standard 
big bang cosmology. Employing standard methods 
\cite{hinova, plin, review}, we can derive the relevant 
condition:
\begin{equation}  \label{Nhi}
\Nhi \equiv \int_{\sigma_{\rm f}}^{\sigma_{*}}\,
\frac{d\sigma}{m^2_{\rm P}}\: \frac{V_{\rm HI}}{V'_{\rm HI}}
\simeq19.4+\frac{2}{3}\ln\frac{V^{1/4}_{\rm HI0}}
{{1~{\rm GeV}}}+\frac{1}{3}\ln\frac{T_{\rm rh}}{{1~{\rm GeV}}},
\end{equation}
where $\sigma_{\rm f}$ is the value of $\sigma$ at the end of FHI, 
$\sigma_{*}$ is the value of $\sigma$ when the pivot scale $k_*$ 
crosses outside the horizon during FHI, the prime in this section 
denotes derivation with respect to $\sigma$, and $T_{\rm rh}$ is 
the reheat temperature after FHI. The value $\sigma_{\rm f}$ can 
be found, in the slow-roll approximation \cite{review}, from the 
condition
\beqs\bea \label{slow} {\sf max}\{\epsilon(\sigma_{\rm f}),
|\eta(\sigma_{\rm f})|\}=1, \eea
where
\bea \label{epsilon}\epsilon\simeq\frac{m^2_{\rm P}}
{2}\left(\frac{V'_{\rm HI}}
{V_{\rm HI}}\right)^2~~\mbox{and}~~\eta\simeq m^2_{\rm P}
\frac{V''_{\rm HI}}{V_{\rm HI}}, \eea\eeqs
or the saturation of the bound in \Eref{Scr}.
\paragraph{}  The amplitude $A_{\rm s}$ of the power spectrum of 
the curvature perturbation which is generated during FHI and
calculated at $k_{*}$ as a function of $\sgm_*$ is to be 
consistent with the present data \cite{wmap, plin}, i.e.
\begin{equation} \label{Prob}
A_{\rm s}^{1/2}=\frac{1}{2\sqrt{3}\, \pi m^3_{\rm P}}
\,\frac{V_{\rm HI}^{3/2}(\sigma_*)}{|V'_{{\rm HI}}(\sigma_*)|}
\simeq 4.685\times 10^{-5}.
\end{equation}

\paragraph{}  The scalar spectral index $n_{\rm s}$, its running
$\alpha_{\rm s}\equiv{d\ns}/{d\ln k}$, and the scalar-to-tensor
ratio $r$,  which are given by
\beqs\bea \label{nS} && n_{\rm s}=1-6\epsilon_*\ +\ 2\eta_*,
\\
&& \label{aS} \alpha_{\rm s}={2}\left(4\eta_*^2-(n_{\rm
s}-1)^2\right)/3-2\xi_*,~~r=16\epsilon_*,~~~~~~
\eea\eeqs
where $\xi\simeq m_{\rm P}^4~V'_{\rm HI} V'''_{\rm HI}/V^2_{\rm
HI}$ and all variables with the subscript $*$ are evaluated at
$\sigma=\sigma_{*}$, should lie in the following 95$\%$ 
\emph{confidence level} (c.l.) ranges \cite{wmap, plin} based on 
the $\Lambda$CDM model:
\beqs\bea\label{nswmap} &&
\ns=0.9603\pm0.014~\Rightarrow~0.945\lesssim n_{\rm s}
\lesssim 0.975,~~~~~~\\
&&\label{obs3}\as=-0.0134\pm0.018,~~\mbox{and}~~
r<0.11. \label{obs4}\eea\eeqs
Limiting ourselves 
to $a_{\rm s}$'s consistent with the assumptions of the 
power-law $\Lambda$CDM cosmological model, we have to ensure 
that $|a_{\rm s}|$ remains negligible. Since, within the 
cosmological models with running spectral index, $|a_{\rm s}|$'s 
of order 0.01 are encountered \cite{plin,wmap}, we impose the 
following upper bound: 
\beq
|a_{\rm s}|\ll0.01.\label{aswmap}
\eeq

\paragraph{}  The mass $M_{W_{\rm R}}$ of the charged 
${\rm SU(2)_R}$ gauge bosons ($W^\pm_{\rm R}$), which are the 
only $\Gsm$ non-singlet superheavy gauge bosons in our case, 
should take the value dictated by the unification of the MSSM 
gauge coupling constants. Using Ref.~\cite{ax3}, we then 
infer that
\beq \label{Mgut} M_{W_R}=g \sqrt{v^2_\Phi +2v^2_T}\simeq2 \times
10^{16}~\GeV~~\mbox{with}~~g\simeq0.7\eeq
being the value of the unified gauge coupling constant.

\paragraph{} The inflationary potential must be bounded below as 
$|S|\to\infty$ to avoid the possibility of a disastrous runaway 
of the system to infinite values of the inflaton field. This 
requirement also facilitates the possibility that the system may 
eventually undergo an inflationary expansion under generic 
initial conditions.

\paragraph{} The expansion of $\Vhios$ in \Eref{Vol} is expected 
to converge at least up to $\sgm\sim\sgm_*$. This can be ensured if, 
for $\sgm\sim\sgm_*$, each successive term in this expansion (and 
the expansion of $K$ in \Eref{K}) is smaller than the previous one. 
In practice, this objective can be easily accomplished if the $k$'s 
in \Eref{K} are sufficiently small.

%
%

\paragraph{} In our model, we were not able to obtain monotonic 
inflationary potentials. The potentials rather develop a maximum 
and a minimum. So the FHI turns out to be of the hilltop type 
\cite{lofti} with $\sigma$ rolling from the region of the 
maximum of the potential down to smaller values. In this case, a 
mild tuning of the initial conditions is required \cite{gpp} in 
order to obtain acceptable $n_{\rm s}$'s. In particular, the lower 
the $n_{\rm s}$ we want to obtain the closer we must set $\sigma_*$ 
to $\sigma_{\rm max}$, where $\sigma_{\rm max}$ is the value of 
$\sigma$ at which the maximum of $V_{\rm HI}$ lies. To quantify 
the amount of this tuning of the initial conditions, we define 
\cite{gpp} the quantity:
\beq \Dex=\frac{\sigma_{\rm max}-\sigma_*}{\sigma_{\rm max}}.
\label{dms}\eeq
The naturalness of the attainment of the hilltop FHI increases 
with $\Dex$. So we must at least require that $\Dex$ 
is not unnaturally small. Moreover, one should avoid the 
possibility that the system is trapped near the minimum of the 
inflationary potential and, consequently, no FHI takes place. 
Probably an era of eternal inflation prior to FHI could be useful 
\cite{lofti} for solving the naturalness problem of the initial 
conditions for the hilltop FHI.

\subsection{Results}\label{num}

As can be easily seen from the relevant expressions above, our
inflationary model depends on the parameters
$$ \kp,~\kpt,~\ld,~M,~\mt,~\ks,~\kss,~\ksss,~\kst,~\ksv.
$$
The first five of these parameters appear in the superpotential -- 
see \Eref{WH} --, while the others appear in the \Ka\ -- see
\Eref{K}. We concentrate on a realization of FHI which attains 
the fulfillment of \Eref{Mgut}, as suggested first in \cref{rlarge} 
and further exemplified in \cref{hinova}. As a consequence of this
equation, $M$ is fixed as a function of the other superpotential
parameters. In our computation, we use $\kpt$, $\mt$, and $\ld$ as 
input parameters and restrict $\kp$ and $\sigma_*$ so that
Eqs.~(\ref{Nhi}) and (\ref{Prob}) are satisfied. The restrictions
on $\ns$ from \Eref{nswmap} can be met by adjusting conveniently
$\ks$ and $\kss$, whereas the last three parameters of the 
\Ka\ control the boundedness below of $\Vhi$. We take $\ksss=1$, 
$\kst=-1$, and $\ksv=0$ throughout the calculation and verify that 
the values of these quantities play no crucial role in the 
inflationary dynamics. Finally, using Eq.~(\ref{aS}), we extract 
$\alpha_{\rm s}$ and $r$.

The crucial difference between our approach and the one of
Refs.~\cite{gpp,king} is, however, the sign of $\ck=\ks$, which 
here is negative -- cf. Refs.~\cite{rlarge,hinova}. As a 
consequence, the fulfillment of \Eref{nswmap} requires negative 
$\ckk$ and, thus, positive $\kss$ -- see \Eref{c4k}. Note that, 
with this choice of signs, $\as$ is somewhat enhanced. More 
explicitly, the potential $\Vhi$, which is given by 
Eqs.~(\ref{Vcor}), (\ref{Vol}), and (\ref{Vtot}), can be 
approximated as
\bea\label{Vnnm} \nonumber \Vhi&\simeq&V_{\rm HIc}+\,\Vhio\,
\left(1+|\ks|\frac{\sigma^2}{2\mP^2}-\,|\ckk|\frac{\sigma^4}
{4\mP^4} \right.\\ &&\left.-\,|\ckx|\frac{\sigma^6}{8\mP^6}+\,
|c_{8K}|\frac{\sigma^8}{16\mP^8}\right),\eea
where the formula for the potential $V_{\rm HIc}$ should be 
taken from \Eref{Vhica} and the fact that $\ckx$ and $c_{8K}$ 
turn out to be 
positive for the values of the parameters chosen here is taken 
into account. As a consequence, $\Vhi$ unavoidably develops a  
non-monotonic behavior. Employing the expression in \Eref{Vnnm}, 
we can show that $\Vhi$ reaches a local maximum at the value
of the inflaton field
\beqs\beq \label{sigmamax}\sigma_{\rm max}\simeq
\frac{\mP \sqrt{\pi|\ks| + \sqrt{\pi^2\ks^2 + (\kp^2
+3\kpt^2)|\ckk|}}}{\sqrt{2 \pi|\ckk|}}\end{equation} 
and a local minimum at 
\begin{equation}
\label{sigmamin} \sigma_{\rm min}\simeq \mP\frac{\sqrt{3 |\ckx| 
+\sqrt{9 \ckx^2 + 32 |\ckk \ckx|}}}{2 \sqrt{|\ckh|}}\cdot\eeq
\eeqs
In deriving \Eref{sigmamax}, we kept terms until the fourth
power of $\sgm$ in the expansion in the RHS of \Eref{Vnnm}, 
whereas, for \Eref{sigmamin}, we focused on the last three 
terms of this expansion and dropped $V_{\rm HIc}$. This is the 
reason why the RHS of the latter formula is independent of 
$V_{\rm HIc}$ and $\ck$.

\begin{figure}[!t]
\centering\includegraphics[width=60mm,angle=-90]{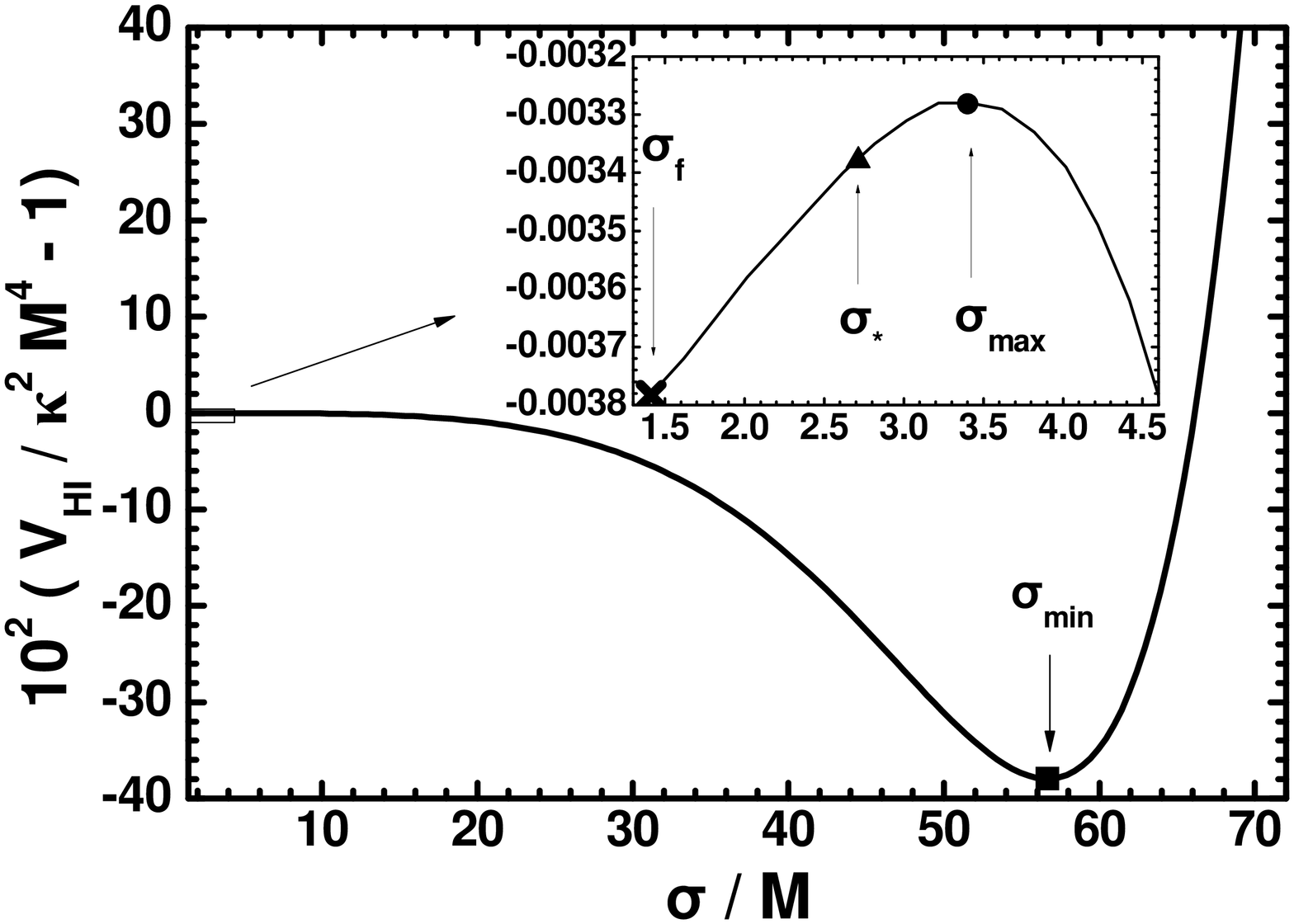}
\caption{\label{sVhi} The variation of $\Vhi$ as a
function of $\sgm$ for $\kp=0.001$, $\kpt=0.01$, $\ld=0.1$, 
$\mt=2.5\times10^{16}~\GeV$, $\ks=-0.0215$, $\kss=10.9$, 
$\ksss=1$, $\kst=-1$, and $\ksv=0$ (resulting to $\ns=0.960$). 
The values $\sigma_*$, $\sigma_{\rm f}$, $\sgm_{\rm max}$, 
and $\sgm_{\rm min}$ of $\sigma$ are also depicted.}
\end{figure}
 
The structure of $V_{\rm HI}$ is visualized in \Fref{sVhi}, 
where we display the variation of $\Vhi$ as a function of 
$\sgm/M$ for $\kp=0.001$, $\kpt=0.01$, $\ld=0.1$, 
$\mt=2.5\times10^{16}~\GeV$, $\ks=-0.0215$, and $\kss=10.9$.
These parameters yield $M\simeq2.6\times10^{16}~\GeV$, $\ns=0.96$,
$\as\simeq0.0013$, and $r\simeq2.25\times10^{-7}$. The maximum of
$\Vhi$ is located at $\sgm_{\rm max}/M=3.4\,\{3.7\}$, whereas its
minimum lies at $\sgm_{\rm min}/M=56\,\{66.5\}$ -- the values
obtained via the approximate \eqs{sigmamax}{sigmamin} are 
indicated in curly brackets. The values of $\sigma_*/M\simeq2.71$ 
and $\sigma_{\rm f}/M\simeq1.41$ are also depicted in the figure. 
The naturalness parameter of the hilltop FHI turns out to be 
$\Dex\simeq0.2$.

\begin{figure*}[!t]
\centering
\includegraphics[width=60mm,angle=-90]{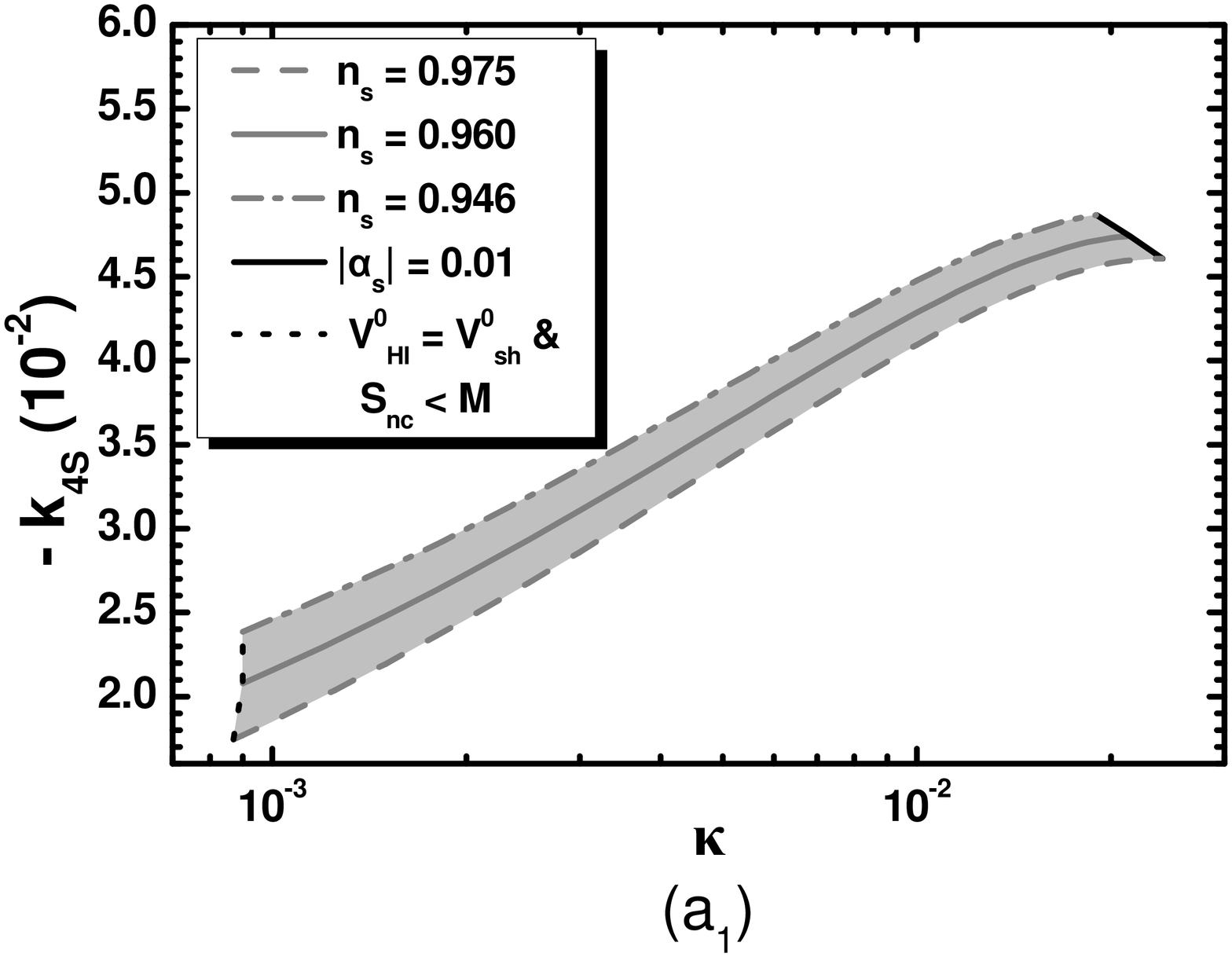}
\includegraphics[width=60mm,angle=-90]{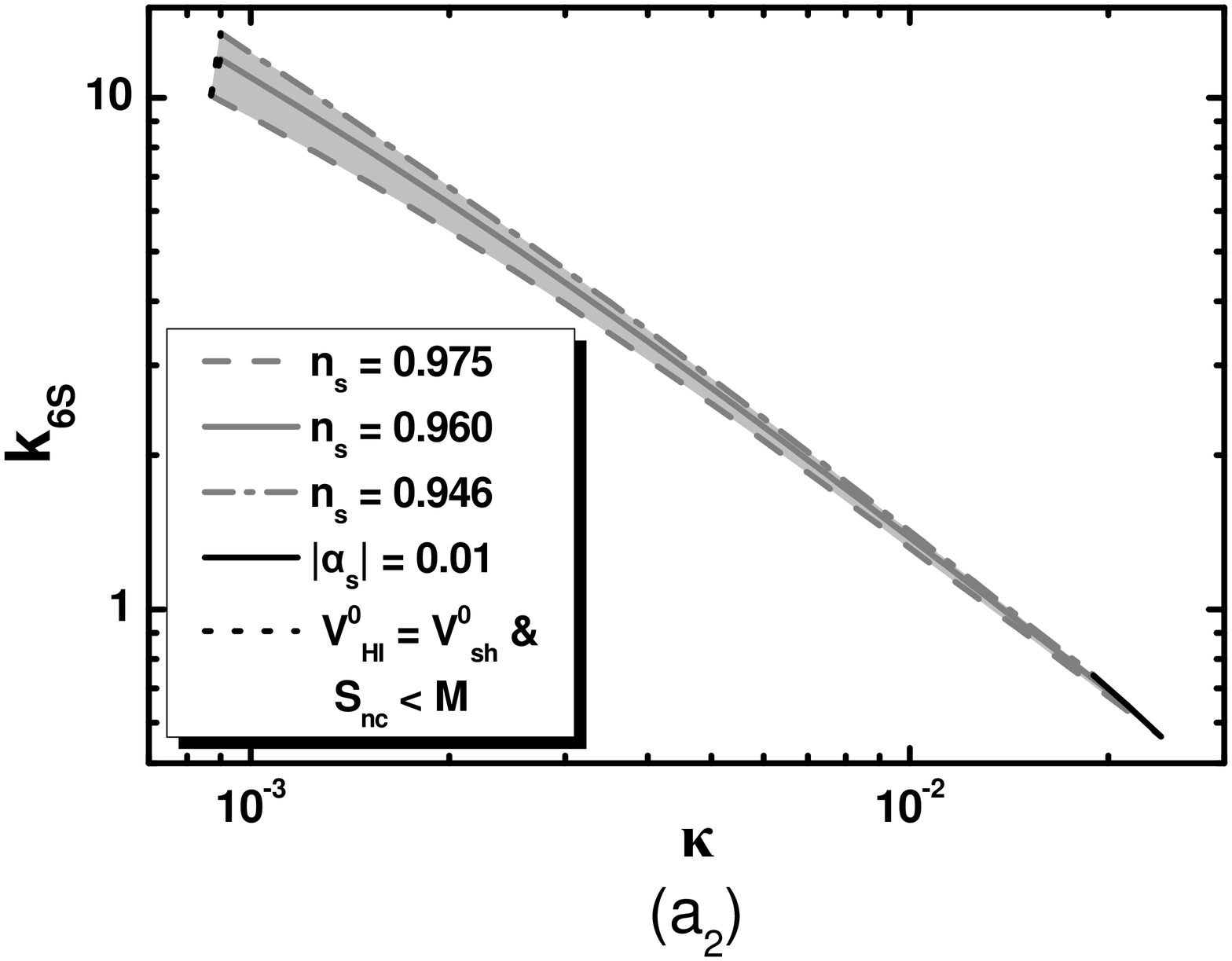}\\
\includegraphics[width=60mm,angle=-90]{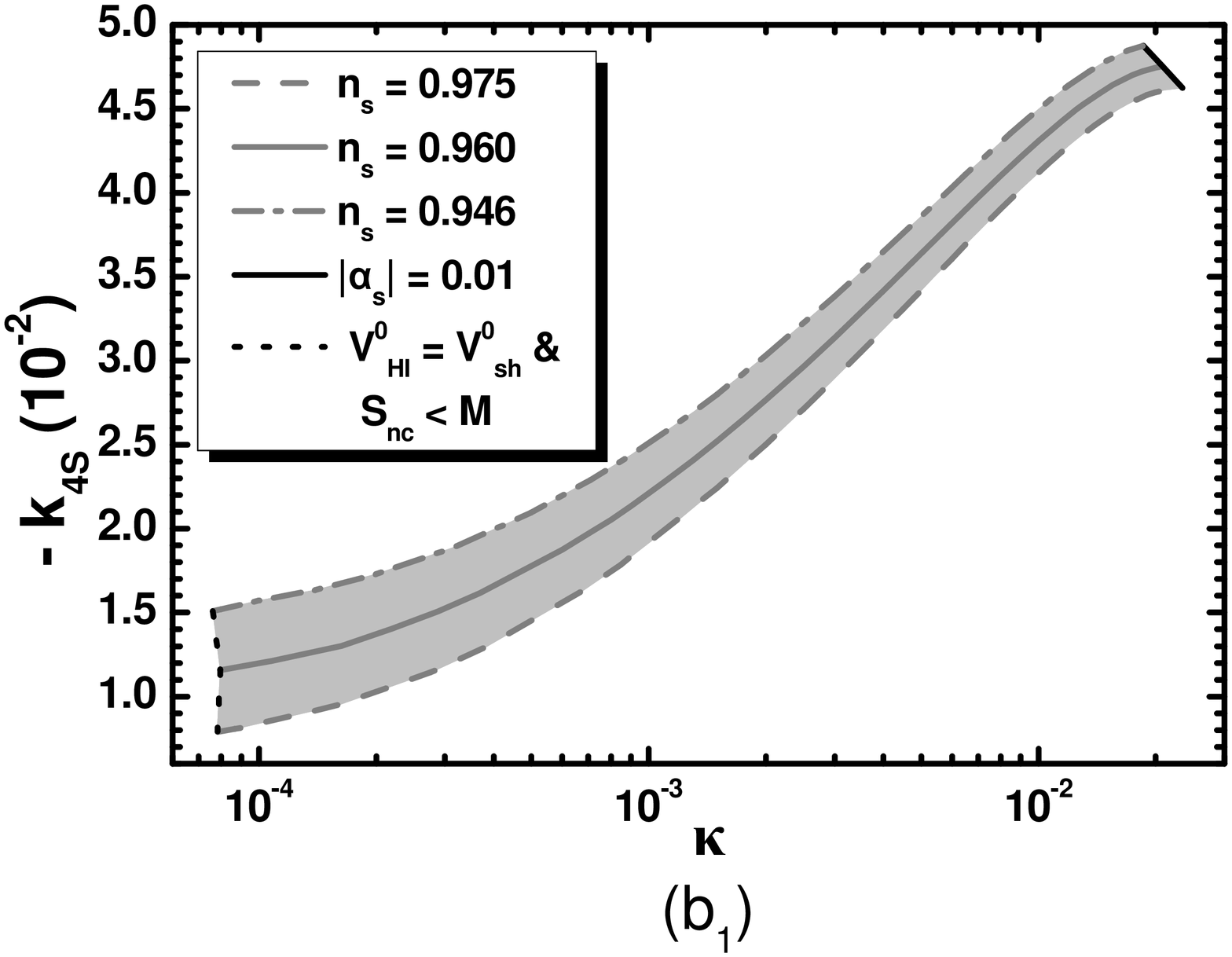}
\includegraphics[width=60mm,angle=-90]{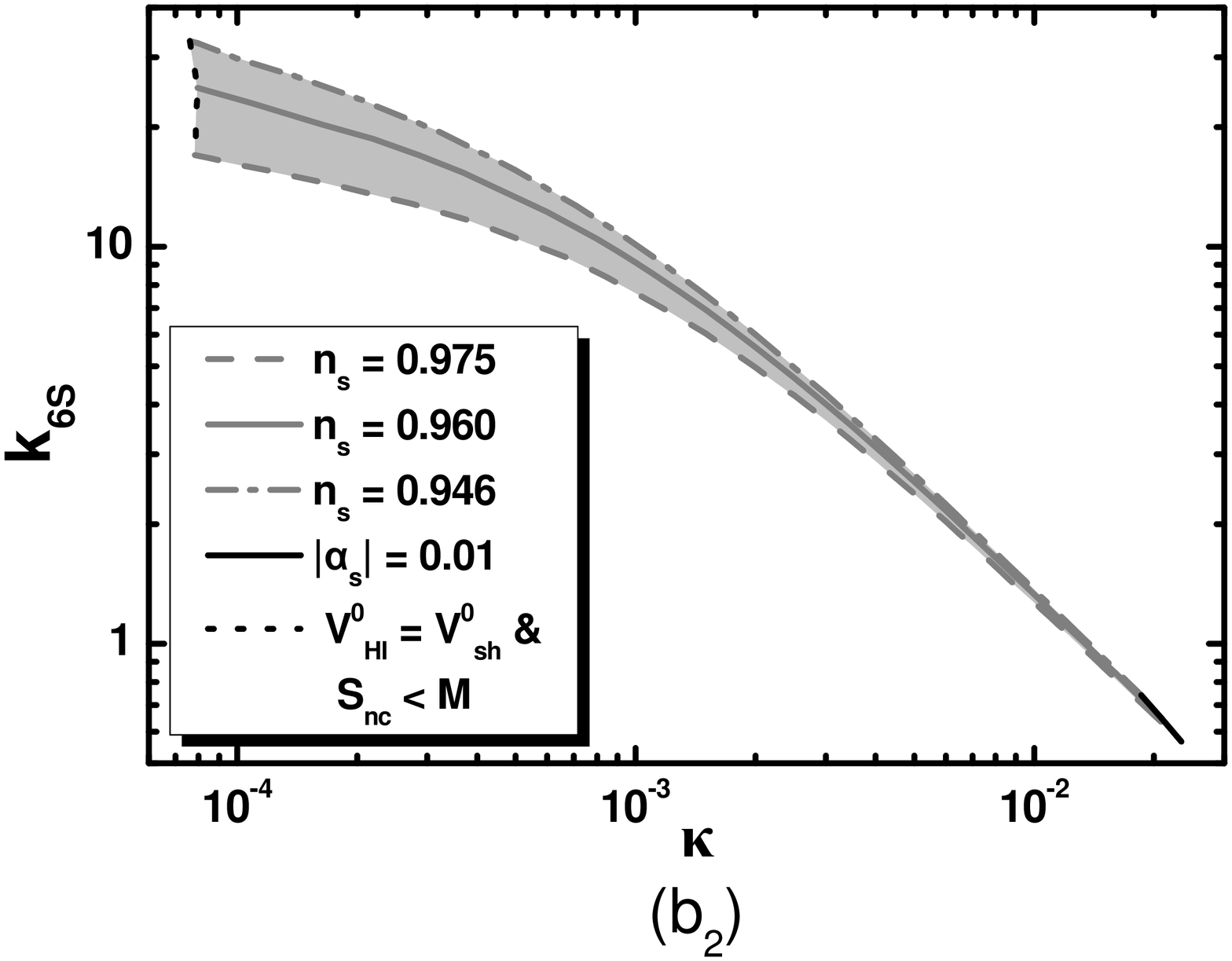}
\caption{\label{fig1}\sl The (shaded) regions allowed by 
Eqs.~(\ref{Vneq1}) or (\ref{Sncr}) as well as Eqs.~(\ref{Nhi}), 
(\ref{Prob}), (\ref{nswmap}), (\ref{aswmap}), and (\ref{Mgut})
in the $\kappa-(-\ks)$ plane (panels ${\rm a}_1$, ${\rm b}_1$) 
and the $\kp-\kss$ plane (panels ${\rm a}_2$, ${\rm b}_2$). We 
take $\ksss=1$ and $\kst=-1$ as well as $\kpt=0.01$, $\ld=0.1$, 
and $\mt=2.5\times10^{16}~\GeV$ for panels ${\rm a}_1$, 
${\rm a}_2$, or  $\kpt=0.005$, $\ld=0.05$, and 
$\mt=3\times 10^{16}~\GeV$ for panels ${\rm b}_1$, ${\rm b}_2$. 
The requirements in the paragraphs e, f, and g of 
Sec.~\ref{cont1} are also satisfied in these regions. The value 
of $n_s$ on the various lines is indicated.}
\end{figure*}

Confronting FHI with the constraints of Sec.~\ref{cont1}, we can
delineate the allowed (lightly gray shaded) region in the
$\kappa-(-\ks)$ [$\kappa-\kss$] plane -- see
Figs.~\ref{fig1}(${\rm a}_1$) and (${\rm b}_1$) 
[Figs.~\ref{fig1}(${\rm a}_2$) and (${\rm b}_2$)]. We take 
$\kpt=0.01$, $\ld=0.1$, and $\mt=2.5\times10^{16}~\GeV$ for panels 
${\rm a}_1$, ${\rm a}_2$ or $\kpt=0.005$, $\ld=0.05$, and 
$\mt=3\times10^{16}~\GeV$ for panels ${\rm b}_1$, ${\rm b}_2$. The 
convention adopted for the various lines is also shown in the 
figure. In particular, the gray dashed [dot-dashed] lines 
correspond to $n_{\rm s}=0.975$ [$n_{\rm s}=0.946$], whereas the 
gray solid lines have been obtained by fixing $n_{\rm s}=0.96$ -- 
see Eq.~(\ref{nswmap}). 

We observe that, as $\kp$ increases, there is a remarkable 
augmentation of $\as$, which saturates the bound in \Eref{aswmap} 
on the thick black solid lines at the right end of the allowed 
regions. The inequalities in \eqs{Vneq1}{Sncr} are violated to the 
left of the black dotted lines. The first of these inequalities, 
though, can remain valid at even smaller values of $\kp$ if we 
take smaller values of $\kpt$ and $\ld$ and larger values of $\mt$ 
and, thus, the dotted line is shifted to the left in this case as 
one can easily deduce by comparing Figs.~\ref{fig1}(${\rm b}_1$) 
and (${\rm b}_2$) with Figs.~\ref{fig1}(${\rm a}_1$) and 
(${\rm a}_2$). This behavior can 
be understood by the fact that, for such values of the parameters, 
the potential $V_{\rm nsh}^0$, which is given by \Eref{Vnsh} -- or 
\Eref{Vnsh1} --, increases and so the bound in \Eref{Vneq1} is 
saturated at smaller values of $\kp$. Note that this bound can 
become totally irrelevant 
for our calculation if we use $\kpt<0$, since, in this case, the 
lower bound on $\xi$ in \Eref{xineq} becomes extremely small and, 
thus, it is automatically satisfied for natural values of $\kp$ 
and $\ld$ 
(of order $0.1$). Would we have used $\kpt<0$ with absolute value 
equal to its values used in \Fref{fig1}, the required values of $\ks$ 
and $\kss$ would have been similar to those found for $\kpt>0$ for 
most of the allowed values of $\kp$ in this figure, but smaller values of 
$\kp$ would also be possible. However, since the achievement of the 
observational constraints of \Sref{cont1} pushes $\kss$ to rather 
high values and $\Dex$ to too small values for such small values of $\kp$, 
it is not worth continuing the exploration of the parameter space in the 
region of such very small $\kp$'s.

Interestingly enough, the allowed regions in 
Fig.~\ref{fig1}(${\rm a}_1$) and (${\rm a}_2$) almost perfectly 
coincide with the allowed regions in
Fig.~\ref{fig1}(${\rm b}_1$) and (${\rm b}_2$) in their common 
range of $\kappa$. This signals the fact that the SUGRA corrections 
to $\Vhi$ originating from the two first terms in the sum in the 
RHS of \Eref{Vol} dominate over the radiative corrections in 
\Eref{Vcor}. The discrepancy between the various lines 
ranges from $2$ to $10\%$. For both sets of values of the input 
parameters, we see that the required values of $|\ks|$ increase 
with $\kp$, whereas the values of $\kss$ drop. Also the mass scale
$M$ increases with $\kp$ and $\mt$. 
As we show in Sec.~\ref{leptR}, $\kp$'s lower than about $0.001$ 
are more preferable from the point of view of non-thermal 
leptogenesis and the $\Gr$ constraint. Focusing on the values of 
the input parameters used in Fig.~\ref{fig1}(${\rm b}_1$) and 
(${\rm b}_2$), which ensure a broader allowed space, and taking 
$\ns\simeq0.96$, we find
\beqs\bea \label{rennm1} &&
0.008\lesssim\frac{\kp}{10^{-2}}\lesssim2.1,
~2.64\lesssim\frac{M}{10^{16}}\lesssim2.85, \\
\label{rennm2} &&  1.15\lesssim\frac{-\ks}{10^{-2}}\lesssim4.7,~~
0.65~\lesssim {\kss}\lesssim25,\\\label{rennm3} &&0.014\lesssim
\frac{-\as}{10^{-2}}\lesssim1,~~2.7\times10^{-5}\lesssim\frac{r}
{10^{-4}}\lesssim2.5.~~~~~~~~\eea\eeqs
In this region, the naturalness parameter $\Dex$ of the hilltop FHI 
ranges between $0.05$ and $0.29$. From the data used in \Fref{fig1}, 
one sees that $\Dex$ increases with $\kp$. These ranges of parameters 
can be further restricted imposing a number of post-inflationary 
requirements as we will see in \Sref{leptR}.

\section{Non-Thermal Leptogenesis}\label{lepto}

In this section, we discuss the inflaton decay and the reheating
of the universe after inflation (\Sref{lept}). We also describe 
the scenario for generating the observed BAU in our model via 
a primordial non-thermal leptogenesis (\Sref{lept1}) consistently 
with the gravitino ($\Gr$) constraint \cite{gravitino,kohri} and 
the low energy neutrino data \cite{forero,fogli} (\Sref{lept2}).

\subsection{The Decay of the Inflaton}\label{lept}

Right after the termination of \FHI, the inflaton field $S$ crosses 
$S_{\rm c}$, the trivial inflationary path in \Eref{flatr} is 
destabilized in the $(\Phi+\bar\Phi^*)/\sqrt{2}$ direction and the 
system is driven towards the SUSY vacuum in \Eref{vev1}. Soon 
afterwards, the system settles into a phase of damped oscillations 
about the SUSY vacuum and eventually decays reheating the universe. 
The constitution of the oscillating \emph{inflaton system} (IS) can 
be found by constructing the neutral scalar particle spectrum at the 
SUSY vacuum in \Eref{vev1}. To this end, we expand $V_{\rm H}$ in
\Eref{potential} up to terms of quadratic order in the fluctuations 
of the fields about the vacuum and find that
\bea V_{\rm H}&\simeq&\lin{\dPp^*}{\dT^*}
M^2_1\stl{\dPp}{\dT}\nonumber\\&&+\lin{\dTb^*}{\dS^*}
M^2_2\stl{\dTb}{\dS}+\cdots, \label{Vexp}\eea
where the (complex) deviations of the fields $S$, $\Phi$, $\bar\Phi$,
$T$, and $\bar T$ from their values in the vacuum are denoted as 
$\dS$, $\dP$, $\dPb$, $\dT$, and $\dTb$ respectively and we have 
defined the complex scalar fields
\beq\dP_{\pm}=\lf\dP\pm\dPb\rg/\sqrt{2}.\eeq
Note that the combination $\dP_{-}$ does not acquire mass from
$V_{\rm H}$ in \Eref{potential} as it is the Goldstone boson absorbed
by the supermassive neutral gauge boson of the model. Recall that 
these complex scalar fields belong to the SM singlet components of 
the various superfields. The mass-squared matrices $M_1^2$ and $M_2^2$ 
in \Eref{Vexp} are given by
\beqs\beq M_1^2=\mtt{2 (\kp^2 + \ld^2) \vp^2}{D_1}{D_1}{\mt^2 + 4
\kpt^2 \vT^2}\eeq with \beq D_1=\sqrt{2} (\ld \mt - 2 \kp \kpt
\vT) \vp\eeq and \beq M_2^2=\mtt{\mt^2 + 2 \ld^2
\vp^2}{D_2}{D_2}{4 \kpt^2 \vT^2 + 2 \kappa^2 \vp^2}\eeq with \beq
D_2=-2 \kpt \mt \vT + 2 \kappa \ld \vp^2.\eeq\eeqs

To find the mass eigenstates of the IS, we have to diagonalize the
matrices above. As it turns out, these matrices have the same 
eigenvalues. So, the diagonalization can be achieved via two 
orthogonal matrices $U_{1,2}$ as follows:
\beq U_1 M_1^2 U_1^\tr =U_2 M_2^2 U_2^\tr =
\mbox{diag}\lf\msp^2,\msm^2 \rg, \eeq where
\beqs\beq \label{mspm} m^2_{\rm I\pm}=\lf\bar m^2\pm D\rg/2\eeq
with
\bea \label{barm}\bar m^2&=&\mt^2 + 4 \kpt^2 \vT^2 + 2 (\kp^2+
\ld^2 )\vp^2,\\ D^2 &=& \bar m^4-8 (\kp \mt + 2 \kpt \ld \vT)^2\vp^2.
\label{Dsq}\eea \eeqs 

The matrices which diagonalize $M^2_1$ and $M^2_2$ can be cast in
the form \beqs\beq
U_n=\mtt{V_{n+}/N_{n+}}{1/N_{n+}}{V_{n-}/N_{n-}}{1/N_{n-}}~~
\mbox{with}~~n=1,2,\eeq
where \beq V_{n\pm}=\frac{C_n\pm D}{2D_n}~~\mbox{and}~~
N_{n\pm}=\sqrt{1+V^2_{n\pm}}.\eeq
Here we use the abbreviations
 \bea C_1 &=& -\mt^2 - 4 \kpt^2 \vT^2 + 2 (\kp^2+\ld^2)\vp^2,
\\ C_2& =& \mt^2 - 4 \kpt^2 \vT^2 - 2 (\kp^2-\ld^2) \vp^2.\eea\eeqs
One can show that $D^2=4D_n^2+C_n^2$ for $n=1$, $2$, which implies
that $D^2$ is positive and, thus, $D$ in Eq.~(\ref{mspm}) is a real 
number taken positive. Also, it is evident that the second term in 
RHS of Eq.~(\ref{Dsq}) is negative and, thus, the masses-squared in 
Eq.~(\ref{mspm}) are both positive.  

Inserting unity ($1=U_n U_n^\tr= U_n^\tr U_n$) on both sides of 
$M^2_1$ and $M_2^2$ in Eq.~(\ref{Vexp}), the potential $V_{\rm H}$ 
can be brought into the form
\beq V_{\rm H}\simeq\sum_{r=\pm}m_{{\rm
I}r}^2\lf|\Phi_{r}|^2+|S_{r}|^2\rg+\cdots\label{Vexp1},\eeq
where the complex fields $\Phi_{\pm}$ and $S_{\pm}$ are given by
\beq \label{SPpm}
\Phi_{\pm}=\frac{\dT+V_{1\pm}\dPp}{N_{1\pm}}~~\mbox{and}~~
S_{\pm}=\frac{\dS+V_{2\pm}\dTb}{N_{2\pm}}.\eeq
Solving Eq.~(\ref{SPpm}) with respect to $\dPp$, $\dT$, $\dTb$, 
and $\dS$, we find
\beqs\bea \label{dPT1}
\dPp&=&\frac{N_{1-}\Phi_{-}-N_{1+}\Phi_{+}}{V_{1-}-V_{1+}},~\\
 \label{dPT2}
 \dT&=&\frac{-N_{1-}V_{1+}\Phi_{-}+N_{1+}V_{1-}
 \Phi_{+}}{V_{1-}-V_{1+}}\eea\eeqs
\mbox{and}~\beqs\bea \label{dSTb1}
\dTb&=&\frac{N_{2-}S_{-}-N_{2+}S_{+}}{V_{2-}-V_{2+}},\\
\label{dSTb2}
\dS&=&\frac{-N_{2-}V_{2+}S_{-}+N_{2+}V_{2-}S_{+}}{V_{2-}-V_{2+}}\cdot
\eea\eeqs

After the end of FHI, each of the four complex scalar fields 
$\Phi_{\pm}$ and $S_{\pm}$ oscillates about the SUSY vacuum 
and decays into a pair of right-handed sneutrinos ($\nu^c_i$) or 
neutrinos ($\psi_{\nu^c_i}$). The masses of these (s)neutrinos are
generated, after the breaking of $G_{\rm LR}$, by the first term
in the RHS of \Eref{Wnr} and turn out to be
\beq \label{rhnmasses} \mrh[i]=2{\ld_{i\nu^c}\vp^2/\Ms}.\eeq
Here we assumed that the superfields $l^c_i$ have been rotated in 
the family space so that the coupling constant matrix $\lambda_{ij}$ 
in \Eref{Wnr} becomes diagonal with real and positive eigenvalues 
$\lambda_{i\nu^c}$. This is the so-called \cite{dent} 
\emph{right-handed neutrino basis}, where the right-handed neutrino 
masses are diagonal, real, and positive. The first coupling in the 
RHS of \Eref{Wnr} 
together with the superpotential terms in \Eref{WH} also leads to 
the decay of the IS to a pair of right-handed neutrinos or 
sneutrinos. In particular, from this coupling, we obtain the 
following Lagrangian term (note that the decay of $T$ via the two 
last terms in the RHS of \Eref{Wm}  is kinematically blocked):
\beqs\bea \nonumber {\cal L}_{\Phi T} &=& - \sqrt{2}
\ld_{i\nu^c}\frac{\vp}{\Ms}\dPp \psi_{\nu_i^c}\psi_{\nu_i^c} +
{\rm H.c.} \\ &=& -\ld_i\sum_{r=\pm}\gamma_{\Phi r}
\Phi_r\psi_{\nu_i^c}\psi_{\nu_i^c} + {\rm H.c.},\label{Lphi} \eea
where \beq \ld_i=\sqrt{2}\ld_{i\nu^c}\vp/\Ms\eeq and 
\beq \label{gP} \gamma_{\Phi r}=
\left\{\bem
- & N_{1+}/(V_{1-}-V_{1+})~~\mbox{for} ~~& r=+ \cr
& N_{1-}/(V_{1-}-V_{1+})~~\mbox{for}~~ & r=-~, \cr 
\eem\right.\eeq\eeqs
as one finds using Eq.~(\ref{dPT1}).

Moreover, from the F-term $(\partial W_{\rm H}/\partial\bar\Phi)^*
(\partial W_{\rm NR}/\partial\bar\Phi)+{\rm H.c.}$ with $W_{\rm H}$ 
and $W_{\rm NR}$ in \eqs{WH}{Wnr} respectively, we obtain the 
Lagrangian terms
\beqs\bea \nonumber {\cal L}_{S\bar T} &=& - 2\vp\ld_{i\nu^c}
\frac{\vp}{\Ms} \lf\kp
S^*+\ld \bar T^*\rg \nu^c_i \nu^c_i + {\rm H.c.} \\
&=& -\ld_i\sum_{r=\pm}\gamma_{Sr} S_rm_{{\rm I}r} \nu^c_i \nu^c_i
+ {\rm H.c.}, \label{LSTb} \eea
where the $\gamma_{Sr}$'s can be derived from \eqs{dSTb1}{dSTb2}
and turn out to be 
\bea\label{gS1}  \gamma_{S +}&=&\frac{\sqrt{2}\vp\lf\kp
N_{2+}V_{2-}-\ld N_{2+}\rg}{m_{{\rm I}+}\lf V_{2-}-V_{2+}\rg},\\
\label{gS2} \gamma_{S -}&=&\frac{\sqrt{2}\vp\lf -\kp
N_{2-}V_{2+}+\ld N_{2-}\rg}{m_{{\rm I}-}\lf V_{2-}-V_{2+}\rg}.\eea
\eeqs

For $ m_{{\rm I}\pm}\gg\mrh[i]$, the Lagrangians ${\cal L}_{\Phi T}$ 
and ${\cal L}_{S\bar T}$ in \eqs{Lphi}{LSTb} give rise to a common 
decay width $\Gm[\rm I +\to\rhni]$ for $\Phi_{+}$ to a pair of 
right-handed neutrinos $\psi_{\nu_i^c}$ and $S_{+}$ to a pair of 
right-handed sneutrinos $\nu^c_i$ and a different common decay width 
$\Gm[\rm I -\to\rhni]$ for $\Phi_{-}$  to a pair of right-handed 
neutrinos $\psi_{\nu_i^c}$ and $S_{-}$ to a pair of right-handed 
sneutrinos $\nu^c_i$:
\beq \Gm[\rm I \pm\to\rhni]\simeq\frac{1}{32\pi}\lambda_i^2\,
\gamma_{\Phi \pm}^2 m_{{\rm I}\pm}=\frac{1}{32\pi}\lambda_i^2\,
\gamma_{S\pm}^2 m_{{\rm I}\pm}. \label{GP}\eeq
The inflaton subsystem consisting of $\Phi_{+}$ and $S_{+}$ will
be called the I$_+$ subsystem, while the one consisting of 
$\Phi_{-}$ and $S_{-}$ will be called the I$_-$ subsystem.
We checked numerically that the widths of the SUGRA-induced 
\cite{Idecay} decay channels of the IS are negligible in our model 
for the values 
of $\vp$ and $\mqn$ obtained in \Sref{num} and, therefore, we do not 
include these channels in our calculation. Since the decay width of 
the produced $\nu^c_{i}$ is much larger than $\Gm[{\rm I}\pm\to\rhni]$ 
-- see below -- the reheating temperature $\Trh$ is exclusively 
determined by the decay of the IS and is given by \cite{quin}
\beq \label{Trh} \Trh=
\left(\frac{72}{5\pi^2g_{*}}\right)^{1/4}\sqrt{\mP\Gm[{\rm
I}-]},~~\mbox{where}~~\Gm[{\rm I}\pm]=\mbox{$\sum_i$}\Gm[\rm I
\pm\to\rhni].\eeq
Here $g_{*}$ counts the effective number of relativistic degrees 
of freedom at temperature $\Trh$ and we assumed that 
$\Gm[{\rm I}-]\ll\Gm[{\rm I}+]$. For the MSSM spectrum plus the 
particle content of the superfields $N$ and $\bar N$, we find
that $g_{*}\simeq228.75+4(1+7/8)=236.25$.

\subsection{Lepton Asymmetry and Gravitino Abundance}
\label{lept1}

The implementation of non-thermal leptogenesis requires that the 
right-handed (s)neutrinos which emerge at reheating decay 
out-of-equilibrium \cite{baryo} to light particles. This 
condition is automatically satisfied provided that 
$\Trh\ll\mrh[i]$. The superfield $\nu^c_{i}$ decays into a 
right-handed Higgs superfield and a ${\rm SU(2)_L}$ doublet 
right-handed antilepton superfield via the tree-level Yukawa 
couplings derived from the second term in the RHS of \Eref{Wy}. 
Interference between tree-level and one-loop diagrams generates 
a lepton-number asymmetry $\ve_i$ per $\nu^c_i$ decay \cite{baryo} 
provided that CP is violated. The resulting overall lepton-number 
asymmetry $Y_L\equiv n_L/s$ ($n_L$ is the lepton-number density 
and $s$ the entropy density) after reheating is given by  
\beqs\beq Y_L=2\frac{5}{4}
\frac{\Trh}{\mqn}\mbox{$\sum_i$}\frac{\Gm[{\rm I-}\to
\nu^c_i]}{\Gqn}\ve_i\label{Yl}\eeq 
and can be partially converted via electroweak sphaleron effects 
into baryon-number asymmetry which, in MSSM, is estimated to be
\beq Y_B=-0.35Y_L.\label{Yb}\eeq\eeqs
The factor 2 in the RHS of \Eref{Yl} comes from the fact that 
each decaying inflaton gives two right-handed (s)neutrinos, 
whereas the factor ($5/4$) is consistent with the calculation 
of $\Trh$ in Ref.~\cite{quin}, which leads to \Eref{Trh}. 
Finally, the numerical factor in the RHS of \Eref{Yb} 
originates \cite{ibanez} from the electroweak sphaleron 
effects.

We should, however, keep in mind that, if the lightest 
right-handed neutrino mass $M_{1\nu^c}$ is less than 
about $10\Trh$, $Y_L$ can be partly washed out due to 
$\nu^c_1$ mediated inverse decay and $\Delta L=1$ 
scattering processes -- this possibility is analyzed in 
\cref{senoguz}. In order to avoid the computational 
complications related to this washout, we limit ourselves 
to cases with $\mrh[1]\gtrsim10\Trh$ so that no washout of 
the non-thermally produced $Y_L$ occurs. Moreover, $Y_L$ is 
not erased by $\Delta L=2$ scattering processes \cite{erasure} 
at all temperatures $T$  between $100~\GeV$ and $\Trh$ since 
$Y_L$ is automatically protected by SUSY \cite{ibanez} for 
$10^7~\GeV\lesssim T\lesssim\Trh$ and for $T\lesssim10^7~\GeV$ 
these processes are well out of equilibrium provided that the
mass of the heaviest light neutrino is smaller than about 
$10~\eV$. This constraint, however, is overshadowed by a more 
stringent restriction induced by the current data 
\cite{wmap,plcp} -- see \Sref{cont2}.

The reheat temperature $\Trh$ must be compatible with the
constraint on the $\Gr$ abundance $Y_{\Gr}$ at the onset of
\emph{big bang nucleosynthesis} (BBN). This abundance is 
estimated to be \cite{kohri}
\beq\label{Ygr} Y_{\Gr}\simeq 1.9\cdot10^{-22}~\Trh/\GeV,\eeq
where we assume that $\Gr$ is much heavier than the gauginos. 
Note that non-thermal $\Gr$ production is \cite{Idecay} also 
possible within SUGRA. However, we adopt here the conservative 
estimate of $Y_{\Gr}$ in \Eref{Ygr} since this non-thermal 
production of gravitinos depends on the mechanism of SUSY 
breaking. It is important to mention that \eqs{Yb}{Ygr} give 
the correct values of baryon asymmetry and $\Gr$ abundance 
provided that no entropy production occurs at $T<\Trh$. This 
requirement can be very easily achieved within our setting.

The mass spectrum of the $N$-$\bar N$ system -- see second 
term in \Eref{Wnr} -- consists of a saxion and an axion 
corresponding, respectively, to the real and the imaginary 
part of the complex scalar field $N_-=(\delta N-\delta\bar N)/
\sqrt{2}$, an axino $\psi_-=(\psi_N-\psi_{\bar N})/\sqrt{2}$, 
two extra real Higgs fields corresponding to 
the real and the imaginary part of $N_+=(\delta N+\delta\bar N)/
\sqrt{2}$, and an extra Higgsino $\psi_{N+}=(\psi_N+\psi_{\bar N})/
\sqrt{2}$ all with masses of order $1~\TeV$ except, of course, 
the axion which is very light ($\delta N$, $\delta\bar N$ are, 
respectively, the complex deviations of $N$, $\bar N$ from 
their VEVs and $\psi$ denotes a Weyl spinor). 

The extra Higgs fields and the extra Higgsino can decay, if 
this is kinematically allowed, to ordinary Higgs fields and 
Higgsinos before dominating the universe \cite{curvaton}. 
However, under certain conditions, the extra Higgsino can 
contribute to the \emph{cold dark matter} (CDM) in the 
universe \cite{antonio}. 

Regarding the saxion 
in $N_{-}$, we can assume that its decay mode to axions is 
suppressed with respect to its decay modes to gluons, 
Higgses, and Higgsinos \cite{Baer, senami} and the initial 
amplitude of its oscillations is approximately equal to the 
axion decay constant $f_a\simeq10^{12}~\GeV$. Under these 
circumstances, the saxion can \cite{Baer} decay before 
dominating the universe and the stringent upper bound on 
$\Trh$ from the limit on the effective number of neutrinos 
at BBN is alleviated \cite{senami}. As a consequence of the 
relatively large decay temperature of the saxion, the 
resulting \emph{lightest sparticles} (LSPs) are likely to be 
thermalized and, therefore, no upper bound on the saxion 
abundance and, thus, $\Trh$ is obtained \cite{senami}. 

The axions could in principle contribute to dark matter, 
but we should keep in mind that they generate isocurvature 
perturbations -- see e.g. Refs.~\cite{curvaton,lchios} -- 
which are strongly restricted by the present data from the 
Planck satellite \cite{plin}. Indeed, since, in our model, 
the PQ symmetry must be broken during FHI -- see 
\cref{curvaton} --, 
the axion acquires quantum fluctuations as all the almost 
massless degrees of freedom. At the QCD phase transition, these 
fluctuations turn into isocurvature perturbations in the axion 
energy density, which means that the partial curvature 
perturbation in axions is different than the one in photons. 
Therefore, a large axion contribution to CDM is disfavored 
within our model. 

Finally, the axino cannot be the LSP because 
its large expected mass and the relatively high $\Trh$'s 
encountered here would then lead \cite{bae} to an unacceptably 
large CDM abundance. Nonetheless, the axino may \cite{bae} 
enhance non-thermally the abundance of a neutralino LSP which 
is a successful CDM candidate.

\subsection{Leptogenesis and Low Energy Neutrino Data} 
\label{lept2}

As mentioned above, the decay of a right-handed sneutrino 
$\nu^c_i$ or neutrino $\psi_{\nu^c_i}$ emerging from the IS 
decay at reheating can generate a lepton asymmetry $\ve_i$ 
due to the interference between the tree-level and the 
one-loop decay diagrams as well as the violation of the CP
symmetry. The generated $\ve_i$ can be expressed in terms of 
the Dirac mass matrix $m^{\rm D}_\nu$ of the neutrinos defined 
in the right-handed neutrino basis:
\beqs\beq\ve_i =\sum_{j\neq i}\frac{\im\left[
(m_{\nu}^{{\rm D}\dag}m^{\rm D}_\nu)_{ij}^2\right]}{8\pi\vev{H_2}^2
(m_\nu^{{\rm D}\dag}m_\nu^{\rm D})_{ii}}
\bigg(F_{\rm V}(x_{ij})+F_{\rm S}(x_{ij})\bigg) ,
\label{el}\eeq 
where 
\beq
x_{ij}\equiv\frac{\mrh[j]}{\mrh[i]} \eeq 
and $\vev{H_2}\simeq174~\GeV$ assuming large $\tan\beta$. 
Also $F_{\rm V}$ and $F_{\rm S}$ represent, respectively, the 
contributions from the vertex and self-energy diagrams and, 
in SUSY theories, are given \cite{covi} by 
\bea \label{fv} F_{\rm V}\lf
x\rg&=&-x\ln\lf1+ x^{-2}\rg, \\ F_{\rm S}\lf
x \rg&=&-\frac{2x}{x^2-1}\cdot\label{fs}\eea \eeqs 
Note that Eqs.~(\ref{el}), (\ref{fv}), and (\ref{fs}) hold 
provided that the right-handed neutrinos are far from being 
degenerate, which is true in our case. In particular, for 
strongly hierarchical $\mrh[i]$'s with $x_{ij}\gg1$, $j\neq i$, 
we obtain the well-known approximate result 
\cite{frigerio,senoguz} 
\beq
\label{hieF} F_{\rm V}+F_{\rm S}\simeq-\frac{3}{x_{ij}^2}
\cdot\eeq

The Dirac mass matrix $m_\nu^{\rm D}$ in \Eref{el} is diagonalized 
in the so-called \cite{dent} \emph{weak basis}, in 
which the lepton Yukawa couplings and the ${\rm SU(2)_L}$ 
interactions are diagonal in the generation space. In 
particular, we have
\beq \label{dD} U^\dag m_\nu^{\rm D} U^{c\dag}\equiv d^{\rm D}=
\diag\lf m^{\rm D}_1,m^{\rm D}_2,m^{\rm D}_3\rg,\eeq 
where $m^{\rm D}_1$, $m^{\rm D}_2$, and $m^{\rm D}_3$ are real 
and positive and 
$U$ and $U^c$ are $3\times3$ unitary matrices which relate 
$l_i$ and $\sni$ in the right-handed neutrino basis with 
$l'_i$ and $\nu^{c\prime}_i$ in the weak basis as follows:
\beq l'= l U\>\>\> \mbox{and}\>\>\>\nu^{c\prime}=U^c \nu^c.\eeq
Here, we write the left-handed ${\rm SU(2)_L}$ doublet lepton 
superfields as row 3-vectors in family space and the right-handed 
${\rm SU(2)_L}$ singlet antilepton superfields as column
3-vectors. The matrix $m_\nu^{{\rm D}\dag}m_\nu^{\rm D}$ in 
\Eref{el} then 
becomes a function of $d^{\rm D}$ and $U^c$. Namely, 
\beq m_\nu^{{\rm D}\dag}m_\nu^{\rm D}=U^{c\dag} d^{{\rm D}\dag}
d^{\rm D}U^c. 
\label{mDD}\eeq

The non-thermal leptogenesis scenario depends on the low 
energy neutrino data via the seesaw formula, which gives the 
light-neutrino mass matrix $m_\nu$ in terms of $m^{\rm D}_i$ and
$\mrh[i]$. In the right-handed neutrino basis, the seesaw 
formula becomes
\beqs\beq \label{seesaw} m_\nu= -m^{\rm D}_\nu\ d_{\nu^c}^{-1}\ 
\left(m_\nu^{\rm D}\right)^{\tr},\eeq 
where \beq d_{\nu^c}=
\diag\lf\mrh[1],\mrh[2],\mrh[3]\rg \eeq\eeqs with
$\mrh[1]\leq\mrh[2]\leq\mrh[3]$ real and positive. Solving
\Eref{dD} with respect to $m_\nu^{\rm D}$ and inserting the 
resulting 
expression in \Eref{seesaw}, we find that the light neutrino 
mass matrix in the weak basis is given by
\beq \label{bmn} \bar m_\nu=U^\dag m_\nu U^*=
-d^{\rm D}U^cd_{\nu^c}^{-1}U^{c\tr}d^{\rm D}.\eeq 
This mass matrix can be diagonalized by the unitary 
\emph{Pontecorvo-Maki-Nakagawa-Sakata} (PMNS) matrix $U_\nu$:
\beq
U_\nu^{\tr}\bar m_\nu U_\nu= \diag\lf\mn[1],\mn[2],\mn[3]\rg\
\label{mns1}\eeq 
with $\mn[1]$, $\mn[2]$, and $\mn[3]$ being the real and positive 
light neutrino mass eigenvalues and the PMNS matrix $U_\nu$ 
parametrized as follows:
\beq \label{mns2} U_\nu = \mtn{c_{12}c_{13}}{s_{12}c_{13}}{s_{13}
e^{-i\delta}} {U_{21\nu}}{U_{22\nu}}{s_{23}c_{13}}
{U_{31\nu}}{U_{32\nu}}{c_{23}c_{13}}\cdot {\cal P}. \eeq
Here 
\beqs\bea U_{21\nu}&=&-c_{23}s_{12}-s_{23}c_{12}s_{13} e^{i\delta},\\
U_{22\nu}&=& c_{23}c_{12}-s_{23}s_{12}s_{13} e^{i\delta},\\
U_{31\nu}&=& s_{23}s_{12}-c_{23}c_{12}s_{13} e^{i\delta},\\
U_{32\nu}&=&-s_{23}c_{12}-c_{23}s_{12}s_{13} e^{i\delta},\eea
\eeqs 
where $c_{ij}\equiv\cos \theta_{ij}$, $s_{ij}\equiv\sin \theta_{ij}$ 
with $\theta_{ij}$ being the appropriate mixing angles and $\delta$ 
is the CP-violating Dirac phase. The two CP-violating
Majorana phases $\varphi_1$ and $\varphi_2$ are contained in the
matrix 
\beq {\cal P}=\diag\lf
e^{-i\varphi_1/2},e^{-i\varphi_2/2},1\rg.\eeq

Following a bottom-up approach along the lines of Refs.~\cite{frigerio,
senoguz}, we find $\bar m_\nu$ via \Eref{mns1} using as input
parameters the low energy neutrino observables for various values of 
$m_{1\nu}$ and the CP-violating Majorana phases $\varphi_1$ and 
$\varphi_2$ and adopting the normal 
or inverted hierarchical scheme of light neutrino masses -- see 
\Sref{leptP}. Taking also $m^{\rm D}_i$ as input parameters, we 
construct the complex symmetric matrix 
\beq
W=-(d^{\rm D})^{-1}\bar m_\nu (d^{\rm D})^{-1}=U^cd_{\nu^c}^{-1}
U^{c\tr}
\label{Wmm}\eeq 
-- see \Eref{bmn} -- from which we can extract $d_{\nu^c}$ as 
follows: 
\beq
d_{\nu^c}^{-2}=U^{c\dag}W W^\dag U^c.\label{WW}\eeq 
Note that $W W^\dag$ is a $3\times3$ complex, Hermitian matrix 
and is diagonalized following the algorithm described in 
\cref{33m} so as to determine the elements of $U^c$ and the 
$\mrh[i]$'s.  We then compute $m_\nu^{{\rm D}\dag}m^{\rm D}_\nu$ 
through \Eref{mDD} and the $\ve_i$'s via \Eref{el}.

\section{Updating the Constraints on the Model Parameters}
\label{cont2}

The parameters of our model can be further restricted if, in
addition to the inflationary requirements mentioned in
\Sref{cont1}, we impose extra constraints arising from the
post-inflationary evolution predicted by our model. These
constraints are outlined in \Sref{leptP}, whereas, in 
\Sref{leptR}, we derive the overall allowed parameter space 
of our model.

\subsection{Post-Inflationary Requirements} \label{leptP}

We summarize below the requirements which guarantee a successful
post-inflationary evolution in our scheme:

\setcounter{paragraph}{0}

\paragraph{} We require the following bounds on $\mrh[i]$:
\beq\label{kin} \mrh[i]\lesssim 7.1\frac{\vp^2}{\Ms},\>\>\mrh[1]
\gtrsim 10\,\Trh,\>\>\mbox{and}\>\>\mqn\geq2\mrh[1].\eeq
The first bound ensures that the coupling constants $\ld_{i\nu^c}$ 
in Eqs.~(\ref{Wnr}) and (\ref{rhnmasses}) acquire perturbative 
values, i.e. $\ld_{i\nu^c}^2/4\pi\leq 1$. The second inequality is 
applied in order to protect the generated lepton asymmetry $Y_L$ 
against any possible washout by $\nu^c_1$-mediated inverse decay 
and $\Delta L=1$ scattering processes as mentioned in 
Sec.~\ref{lept1} -- see Ref.~\cite{senoguz}. Finally, the last 
bound ensures that the decay of the IS to a pair of $\sni$'s is 
kinematically allowed for at least one species of the $\sni$'s.

\paragraph{} The Dirac masses $m^{\rm D}_i$ selected for $\nu_i$ at
\Mgut\ need to be consistent with the relations in
\eqs{mDirac}{mE}. In order to reduce the number of free parameters 
and simplify the relevant constraint, we assume that $y_{ijL}$ and 
$y'_{ijL}$ are simultaneously diagonal in the weak basis with 
elements $y_{iL}$ and $y'_{iL}$ respectively. Under this
assumption, we have to check that the selected $m^{\rm D}_i$'s can be
obtained together with the masses $m_{iE}$ of the charged leptons 
by a natural set of $y_{iL}$'s and $y'_{iL}$'s with $a_1$ and
$a_2$ of order unity. In other words, the solution of the six by
six system of equations
\beq \frac{y_{iL}-\alpha_2y'_{iL}}
{\sqrt{1+|\alpha_2|^2}}\,v_2=m^{\rm D}_i,~~
\frac{y_{iL}-\alpha_1y'_{iL}}
{\sqrt{1+|\alpha_1|^2}}\,v_1=m_{iE}
\label{Ysol}\eeq
has to exist and be natural for a set of natural values of $a_1$ and
$a_2$. Here we put $v_1=174\cos\beta~\GeV$ and $v_2=174\sin\beta~\GeV$, 
and $m_{iE}$ and $m^{\rm D}_i$ are taken at $\Mgut$ assuming that the 
running from $\Mgut$ until the scale of non-thermal leptogenesis 
$\Lambda_L$, which is taken to be $\Lambda_L=\mqn$, is negligible. 
Working in the context of MSSM with universal gaugino masses and 
$\tan\beta\simeq50$ -- favored by the recent results of LHC 
\cite{higgs} on the lightest Higgs boson mass -- and taking into 
account the SUSY threshold corrections, we obtain \cite{fermionM}
\bea\nonumber &&\hspace*{2.1cm}(m_{1E},m_{2E},m_{3E})=\\
&& (0.39-0.532,83.5-112.7,1635-2400)~\MeV. ~~~\label{m123E}\eea

\paragraph{} From the solar, atmospheric, accelerator,
and reactor neutrino experiments, we take as inputs in our 
calculation the best-fit values \cite{forero} -- see also 
\cref{fogli} -- 
\beqs\bea \label{msol} \Delta m^2_{21}&=&7.62\times 10^{-3}~{\rm
eV}^2,\\  \Delta m^2_{31}&=&
2.55\left[-2.43\right]\times 10^{-3}~{\rm eV}^2
\label{matm}\eea
for the differences $\Delta m^2_{ij}\equiv m^2_{i\nu}-m^2_{j\nu}$ 
between the light neutrino masses-squared, 
\bea \sin^2\theta_{12}&=&0.32,\\
\sin^2\theta_{13}&=&0.0246\left[0.025\right],\\
\sin^2\theta_{23}&=&0.613\left[0.6\right] \label{8exp}\eea
for the mixing angles, and
\beq\delta=0.8\pi\left[-0.03\pi\right] \label{dexp}\eeq\eeqs
for the CP-violating Dirac phase in the case of normal 
[inverted] neutrino mass hierarchy. In particular, two of the 
$\mn[i]$'s are determined in terms of the third one using the 
relation
\beqs
\beq\mn[2]=\sqrt{\mn[1]^2+\Delta m^2_{21}}\eeq 
and either
\beq\mn[3]=\sqrt{\mn[1]^2+\Delta m^2_{31}}\eeq 
for \emph{normally ordered} (NO) $\mn[i]$'s 
or 
\beq\mn[1]=\sqrt{\mn[3]^2+\left|\Delta
m^2_{31}\right|}\eeq\eeqs 
for \emph{invertedly ordered} (IO) $\mn[i]$'s. We also take 
into account the fact that the sum of the $\mn[i]$'s is 
bounded above by the current data 
\cite{wmap, plcp}: 
\beq\label{sumn} \mbox{$\sum_i$}\mn[i]\leq0.28~{\eV}\eeq 
at 95\% c.l.

\paragraph{} The BAU $Y_B$ must satisfy the constraint \cite{plcp}
\beq Y_B\simeq\lf8.55\pm0.217\rg\times10^{-11}~~\mbox{at 95\%
c.l.}\label{BAUwmap}\eeq

\paragraph{} To avoid spoiling the success of the BBN, an upper 
bound on $Y_{\Gr}$ must be imposed depending on the $\Gr$ mass 
$m_{\Gr}$ and the dominant $\Gr$ decay mode. We consider 
here the conservative case where $\Gr$ decays with a tiny 
hadronic branching ratio. In this case, we have \cite{kohri}
\beq  \label{Ygw} Y_{\Gr}\lesssim\left\{\bem
%
10^{-14}\hfill \cr
10^{-13}\hfill \cr
10^{-12}\hfill \cr \eem
\right.\>\>\>\mbox{for}\>\>\>m_{\Gr}\simeq\left\{\bem
0.69~{\rm TeV}\hfill \cr
10.6~{\rm TeV}\hfill \cr
13.5~{\rm TeV}.\hfill \cr \eem
\right.\eeq

\subsection{Results}\label{leptR}

The inflationary requirements of \Sref{fhi} restrict $\ks$ and
$\kss$ as functions of $\kp$ for given $\ld$, $\kpt$, and $\mt$. 
We first concentrate on a low value of $\kp$ within its allowed 
range. This ensures a low enough $\mqn$ through \Eref{mspm}. As a
consequence, $Y_B$ in \Eref{Yb} is enhanced, whereas $\Trh$ 
is kept sufficiently low, as can be deduced from \eqs{GP}{Trh}.
Namely, we take $\kp=0.001$, $\kpt=0.01$, $\ld=0.1$, 
$\mt=2.5\times10^{16}~\GeV$, $\ks=-0.0215$, and $\kss=10.9$
yielding $\mqn=2.94\times10^{13}~\GeV$.

Note that $\Trh$ and $Y_B$ depend also on the masses $\mrh[i]$ 
of the $\sni$'s into which ${\rm I}_-$ decays. In addition, 
$Y_B$ depends crucially on the low energy parameters related to 
neutrino physics. Following a bottom-up approach, we find the
$\mrh[i]$'s by using as input parameters the $m^{\rm D}_i$'s, 
the mass of one of the $\nu_i$'s -- the $\mn[1]$ for NO 
$\mn[i]$'s, or the $\mn[3]$ for IO $\mn[i]$'s --, the two 
Majorana phases $\varphi_1$ and $\varphi_2$ of the PMNS matrix, 
and the best-fit values -- see Eqs.~(\ref{msol})-(\ref{dexp}) 
-- of the low energy neutrino parameters. In our numerical code, 
we run these best-fit values up to the scale of non-thermal 
leptogenesis $\Lambda_L=\mqn$ following Ref.~\cite{running} and 
considering the MSSM with $\tan\beta\simeq50$ as an effective 
theory between the soft SUSY-breaking scale 
$M_{\rm SUSY}=1.5~\TeV$ and $\Lambda_L$. The so obtained 
$\mrh[i]$'s clearly correspond to the scale $\Lambda_L$. 

Our results are displayed in \Tref{tab4} for some
representative values of the parameters which yield acceptable
$\Yb$ and $Y_{\Gr}$, i.e. lying in the ranges shown in 
\eqs{BAUwmap}{Ygw}. We consider strongly NO (cases A and B), 
almost degenerate (cases C, D, and E) and strongly IO (cases F 
and G) neutrino masses. Note that the cases C and D correspond to 
NO $\mn[i]$'s with large $\mn[1]$, while the case E corresponds to 
IO $\mn[i]$'s with large $\mn[3]$. In all these cases, the current 
limit -- see \Eref{sumn} -- on the sum of the $\mn[i]$'s is safely 
met -- in the case D, this limit is almost saturated. Care is 
taken, in addition, so that the first inequality of \Eref{kin} is 
satisfied. Our choice to use the effective scale $M_S$ in \Eref{Wnr} 
helps in this direction. Indeed, have we chosen this effective 
scale to be equal to $\mP$, the case A in \Tref{tab4} would be 
excluded due to the violation of this inequality. We also 
observe that with strongly NO or IO $\mn[i]$'s the resulting 
$\mrh[i]$'s are strongly hierarchical. With almost degenerate 
$\mn[i]$'s, though, the resulting $M_{i\nu}$'s are closer to one 
another. As a consequence, in this case, more ${\rm I}_-$-decay 
channels are, generally, available. In the case A, only a single 
decay channel is open. In all the other cases, the dominant 
contribution to $\Yb$ arise from $\ve_2$ -- recall Eqs.~(\ref{Yl})
and (\ref{Yb}). In \Tref{tab4}, we also display, for comparison, 
the $B$ abundance with ($Y_B$) or without ($Y^0_B$) taking into 
account the renormalization group running of the low energy 
neutrino data. We observe that the two results are in most cases 
close to each other with the biggest discrepancy encountered in 
the case E of almost degenerate IO $\mn[i]$'s. Shown are also the 
values of $\Trh$, the majority of which are close to 
$5\times 10^8~\GeV$, and the corresponding $Y_{\Gr}$'s, which, in 
most of the cases, are consistent with \Eref{Ygw} only for large 
values of $m_{\Gr}\gtrsim 10~\TeV$. Thus, from the perspective of 
the $\Gr$ constraint, the case A turns out to be the most promising 
one.

\begin{table}[!t]
\caption{Parameters yielding acceptable BAU for $\kp=0.001$,
$\kpt=0.01$, $\ld=0.1$, $\mt=2.5\times10^{16}~\GeV$, 
$\ks=-0.0215$, $\kss=10.9$, and various neutrino mass schemes.}
\begin{tabular}{c||c|c||c|c|c||c|c}\toprule
Parameters &  \multicolumn{7}{c}{Cases}\\\cline{2-8}
&A&B& C & D& E & F&G\\ \cline{2-8} &\multicolumn{2}{c||}{Normally} &
\multicolumn{3}{|c||}{Almost}&  \multicolumn{2}{|c}{Invertedly}
\\& \multicolumn{2}{c||}{Ordered}&\multicolumn{3}{|c||}{Degenerate}& 
\multicolumn{2}{|c}{Ordered}\\
& \multicolumn{2}{c||}{$\nu$ Masses}&\multicolumn{3}{|c||}{$\nu$ Masses}& 
\multicolumn{2}{|c}{$\nu$ Masses}\\
\colrule
\multicolumn{8}{c}{Low Energy Neutrino Parameters}\\\colrule
$\mn[1]/0.1~\eV$&$0.01$&$0.1$& $0.5$& $0.7$&$0.7$ & $0.5$&$0.49$\\
$\mn[2]/0.1~\eV$&$0.09$&$0.1$& $0.5$& $0.7$&$0.7$ & $0.51$&$0.5$\\
$\mn[3]/0.1~\eV$&$0.5$&$0.5$& $0.7$& $0.86$&$0.5$ &
$0.1$&$0.05$\\\colrule
$\sum_i\mn[i]/0.1~\eV$&$0.6$&$0.7$& $1.7$& $2.3$&$1.9$ & $1.1$&$1$\\
\colrule
$\varphi_1$&$\pi/3$&$\pi/2$& $0$& $\pi/2$&$\pi$&$-\pi/3$ & $-\pi/2$\\
$\varphi_2$&$0$&$0$& $3\pi/4$& $\pi/2$&$\pi$ &
$-\pi/2$&$-\pi/6$\\\colrule
\multicolumn{8}{c}{Mass Parameters at the Leptogenesis Scale }\\\colrule
$m^{\rm D}_1/0.1~\GeV$&$4.7$&$4.1$& $15.5$& $10$&$7$ & $9.5$&$7$\\
$m^{\rm D}_2/\GeV$&$26$&$2.3$& $2$& $2.5$&$1.2$ & $1.4$&$2$\\
$m^{\rm D}_3/10~\GeV$&$12$&$12$& $5$& $8$&$0.4$ & $12$&$1.5$\\\colrule
$\mrh[1]/10^{10}~\GeV$&$5.9$&$2.2$& $4.9$& $1.4$&$0.67$ & $1.7$&$1$\\
$\mrh[2]/10^{11}~\GeV$&$177$&$1.4$& $1$& $0.94$&$0.069$ & $0.8$&$1.5$\\
$\mrh[3]/10^{13}~\GeV$&$342$&$45$& $1.9$& $5.3$&$0.007$ &
$51$&$1.7$\\\colrule
\multicolumn{8}{c}{Decay Channels of the Inflaton ${\rm I}_-$ with mass
$\mqn$}\\\colrule
I$_-\to$&$\wrhn[1]$&$\wrhn[1,2]$& $\wrhn[1,2]$& $\wrhn[1,2]$&
$\wrhn[1,2,3]$ & $\wrhn[1,2]$&$\wrhn[1,2]$\\ \colrule
\multicolumn{8}{c}{Resulting Baryon Asymmetry }\\\colrule
$10^{11}Y^0_B$&$8.72$&$7.45$& $7.98$& $7.96$&$5.5$ & $7.97$&$7.97$\\
$10^{11}Y_B$&$8.53$&$8.23$& $8.4$& $8.64$&$8.78$ &
$8.6$&$8.53$\\\colrule
\multicolumn{8}{c}{Resulting $\Trh$ and $\Gr$ Abundance}\\\colrule
$\Trh/10^{8}~\GeV$&$3.4$&$8$& $6.8$& $5.6$&$5.9$ & $4.9$&$8.7$\\
$10^{13}Y_{\Gr}$&$0.7$&$1.5$& $1.3$& $1$&$1.1$ &
$0.9$&$1.65$\\\botrule
\end{tabular}
\label{tab4}
\end{table}

As we emphasize in \Sref{theory}, the inclusion in our model of the 
$T$ and $\bar T$ superfields -- which has various consequences for 
the inflationary scenario (see \Sref{fhi}) --  is of crucial 
importance for the violation of the partial YU  and the tight 
constraint on the Dirac neutrino masses $m^{\rm D}_i$'s predicted 
by the 
simplest left-right symmetric model. Indeed, in the simplest model, 
where $\alpha_1=\alpha_2$, and for the central values of the 
$m_{iE}$'s in \Eref{m123E}, we would have the following values of 
the $m^{\rm D}_i$'s:
\beq \lf m^{0{\rm D}}_{1},m^{0{\rm D}}_{2},m^{0{\rm D}}_{3}
\rg\simeq (0.023,4.9,100)~\GeV. 
\label{m0D}\eeq
However, in sharp contrast with Eq.~(\ref{m0D}), in all the cases 
presented in \Tref{tab4}, $m^{\rm D}_1\gtrsim 0.1~\GeV$. Such large 
values of $m^{\rm D}_1$ are necessary in order to be able to fulfill
the second inequality in \Eref{kin}, given that $m^{\rm D}_1$ 
heavily influences 
$\mrh[1]$. The extended left-right symmetric model described in 
\Sref{theory} gives us a much larger flexibility in selecting 
appropriate $m^{\rm D}_i$'s with natural values of the Yukawa 
coupling constants and $\alpha_1\neq\alpha_2$ of order unity. To
highlight further this key issue of our work, we display in
\Tref{tab5} solutions to \Eref{Ysol} for the cases displayed in
\Tref{tab4}, central values of the input parameters in \Eref{m123E},
$a_1=1.2$, and $a_2=0.5$. We see that all the Yukawa coupling
constants listed in this table take natural values without any 
ugly hierarchy being necessary in any pair $(y_{iL},y'_{iL})$.

\begin{table}[!t]
\caption{Solutions to Eq.~(\ref{Ysol}) for the 
cases displayed in
\Tref{tab4}, central values of the input parameters in \Eref{m123E},
$a_1=1.2$, and $a_2=0.5$.}
\begin{tabular}{c||c|c||c|c||c|c}\toprule
Case &
$y_{1L}$&$y'_{1L}$&$y_{2L}$&$y'_{2L}$&$y_{3L}$&$y'_{3L}$\\\colrule
A& $0.005$&$0.004$&$0.24$&$0.17$&$0.67$&$-0.19$\\
B& $0.0044$&$0.0034$&$-0.006$&$-0.042$&$0.67$&$-0.19$\\\colrule
C& $0.0017$&$0.0014$&$-0.0094$&$-0.044$&$-0.096$&$-0.83$\\
D& $0.011$&$0.0088$&$-0.0039$&$-0.04$&$0.23$&$-0.56$\\
E& $0.0075$&$0.0061$&$-0.018$&$-0.052$&$-0.6$&$-1.26$\\\colrule
F& $0.01$&$0.008$&$-0.016$&$-0.05$&$0.67$&$-0.19$\\
G& $0.0076$&$0.0061$&$-0.009$&$-0.044$&$-0.48$&$-1.15$\\\botrule
\end{tabular}
\label{tab5}
\end{table}

In order to extend our conclusions inferred from \Tref{tab4} to 
the case of a variable $\kappa$, we
now examine how the central value of $Y_B$ in \Eref{BAUwmap} can 
be achieved by varying one of the $m^{\rm D}_i$'s as a function of
$\kp$ or $\mqn$. To this end, we fix $\ns$ to its central value in
\Eref{nswmap} and $\kpt$, $\ld$, $\mt$, $\ksss$, $\kst$, and $\ksv$ 
to their values corresponding to Figs.~\ref{fig1}(${\rm b}_1$) and 
(${\rm b}_2$). Consequently, the parameters $\ks$ and $\kss$ vary 
with $\kappa$ along the solid 
gray lines in these figures. Moreover, we set the values of the 
$m_{i\nu}$'s (by selecting $\mn[1]$ for NO $\mn[i]$'s or $\mn[3]$ 
for IO $\mn[i]$'s), $m^{\rm D}_1$, $m^{\rm D}_3$, $\varphi_1$, and 
$\varphi_2$ equal to their values in the cases B, D, or F of 
\Tref{tab4}. Since, in these cases, I$_-$ decays mainly into 
$\nu_2^c$ with $\mrh[2]>\mrh[1]$, the value of $\mrh[2]$ heavily 
influences $Y_B$. In turn, the variation of $\mrh[2]$ is almost 
exclusively due to the variation $m^{\rm D}_2$ -- see approximate 
formulas of \cref{senoguz}.

\begin{figure}[!t]
\centering\includegraphics[width=60mm,angle=-90]{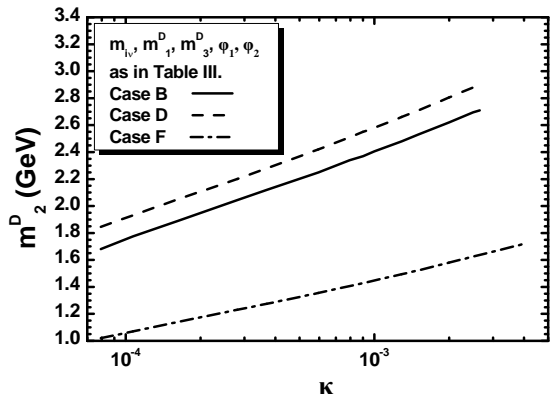}
\caption{\label{kpmD} Contours in the $\kp-m^{\rm D}_2$ plane 
yielding the central $Y_B$ in \Eref{BAUwmap} consistently with 
the inflationary requirements for $\kpt=0.005$, $\ld=0.05$, 
$\mt=3\times10^{16}~\GeV$, $\ksss=1$, $\kst=-1$, $\ksv=0$, 
$\ns=0.96$, and the values of $m_{i\nu}$, $m^{\rm D}_1$, 
$m^{\rm D}_3$, $\varphi_1$, and $\varphi_2$ which correspond 
to the cases B (solid line), D (dashed line), and F (dot-dashed 
line) of \Tref{tab4}.}
\end{figure}

The resulting contours in the $\kp-m^{\rm D}_2$ plane are presented 
in \Fref{kpmD} -- since the range of $Y_B$ in \Eref{BAUwmap} is very 
narrow, the $95\%$ c.l. width of these contours is negligible. The
convention adopted for these lines is also described in the figure. 
In particular, we use solid, dashed, or dot-dashed line for $\mn[i]$, 
$m^{\rm D}_1$, $m^{\rm D}_3$, $\varphi_1$, and $\varphi_2$ 
corresponding to the cases B, D, or F of \Tref{tab4} respectively. 
The lower limit on these lines comes from the violation of
\eqs{Vneq1}{Sncr} -- as in Figs.~\ref{fig1}($b_1$) and ($b_2$). At 
the other end, these lines terminate at the values 
of $m^{\rm D}_2$ beyond which the second inequality in \Eref{kin} is 
violated and, therefore, washout effects start becoming significant. 
At these upper termination points of the contours, we obtain 
$\Trh\simeq2\times10^9~\GeV$ 
or $Y_{\Gr}\simeq4\times10^{-13}$ and so we expect that the constraint 
of \Eref{Ygw} will cut any possible extension of the these curves 
beyond these termination points 
that could survive the possible washout of $Y_L$. Along the depicted 
contours, we obtain $8\times10^{-2}\lesssim\kp/10^{-3}\lesssim4$,
$2.3\lesssim\mqn/10^{12}\GeV\lesssim200$, whereas the naturalness 
parameter of the hilltop FHI $\Dex=0.05-0.27$.
Also the resulting $\mrh[2]$'s vary in the range
$(4-19)\times10^{10}~\GeV$ and $\mrh[1]$ remains close to
$(1-2)\times10^{10}~\GeV$. The values of $y_{2L}$, $y'_{2L}$
selected in \Tref{tab4} for the cases B, D, and F change also 
along the displayed curves of \Fref{kpmD}, without any essential
modification though as regards their general features. 

\section{Conclusions}\label{concl}

We constructed a SUSY GUT model based on the left-right symmetric 
gauge group $\Ggut$, which supports FHI followed by successful 
reheating and non-thermal leptogenesis. The lepton-number asymmetry 
is generated via the decay of the right-handed neutrinos $\nu^c_i$ 
which emerge from the decay of the inflaton system during the 
reheating process. It is important that any possible washout of 
the produced lepton 
asymmetry can be avoided. Our proposal is tied to the addition of 
two pairs of superfields (one pair consisting of bidoublets under 
${\rm SU(2)_{L}}\times {\rm SU(2)_{R}}$ and another consisting of 
triplets under ${\rm SU(2)_{R}}$) -- see Table I --, which 
naturally leads to an adequately strong violation of the 
asymptotic partial YU predicted by the simplest left-right 
symmetric model of \cref{vlachos}. Confining 
our discussion to the trivial inflationary path, we found that 
the extra triplets play a crucial role {\it (i)} in the inflationary 
scenario causing extra radiative corrections along the inflationary 
path, and {\it (ii)} in the reheating process assisting us in 
obtaining an acceptably low reheat temperature. 

We expanded the \Ka\ -- see \Eref{K} -- up to 
twelfth order in powers of the various fields and selected a 
convenient choice of signs which ensures that the parameters of 
the superpotential of our model assume values compatible with the 
requirement of gauge coupling constant unification within MSSM 
with the inflationary potential $\Vhi$ remaining bounded below at 
least up to the Planck scale $\mP$. The FHI reproduces the current 
data on the amplitude $A_{\rm s}$ 
of the power spectrum of the curvature perturbation and the scalar
spectral index $n_{\rm s}$ within the power-law $\Lambda$CDM 
cosmological model and generates the number of e-foldings required 
for the resolution of the horizon and flatness problems of the 
standard big bang cosmological model. 

Imposing additional constraints from the BAU, the (unstable) 
gravitino abundance, and the neutrino oscillation parameters, we 
concluded that, for the central value of $\ns$, $\kp\simeq8
\times 10^{-5}-0.004$ and $m^{\rm D}_1
\gtrsim0.1~\GeV$ with the remaining parameters of the 
superpotential of our model taking more or less natural values, 
whereas the naturalness parameter for the hilltop FHI $\Dex\simeq
0.05-0.27$. It is gratifying that our model exhibits solutions 
with the inflaton system decaying exclusively into the lightest 
of the right-handed neutrinos $\nu^c_i$. These solutions are the 
most promising from the perspective of the gravitino constraint. 

\acknowledgments We would like to thank A. Pilaftsis for an
enlightening correspondence. This work was supported by the 
European Union under the Marie Curie Initial Training Network 
`UNILHC' PITN-GA-2009-237920. The work of R.A. was supported 
by the Tomalla Foundation and C.P. acknowledges support from 
the Generalitat Valenciana under grant PROMETEOII/2013/017.

\appendix

\renewenvironment{subequations}{%
\refstepcounter{equation}%
\setcounter{parentequation}{\value{equation}}%
  \setcounter{equation}{0}
  \def\theequation{A\theparentequation{\sffamily\alph{equation}}}%
  \ignorespaces
}{%
  \setcounter{equation}{\value{parentequation}}%
  \ignorespacesafterend
}

\section{Reheating Process, Lepton Asymmetry and Gravitino 
Abundance}
\label{Rhg}

In this Appendix, we present a numerical description of the
post-inflationary evolution of the various energy and number
densities involved in our scenario of non-thermal leptogenesis.

In particular, the energy densities $\rho_+$ and $\rho_-$ of the
${\rm I}_+$ and ${\rm I}_-$ subsystems respectively -- see the
definition of these subsystems right after \Eref{GP} --, the 
energy density 
$\rho_{\rm R}$ of the produced radiation, and the number densities 
$n_L$ of the leptons and $n_{\Gr}$ of the $\Gr$'s satisfy the 
following Boltzmann equations -- cf. Refs.~\cite{gpp, kohri}:
\beqs\begin{eqnarray} && \dot \rho_++3H\rho_++\Gamma_{\rm I+}
\rho_+=0,\label{nf}\\
&& \dot\rho_-+3H\rho_-+\Gamma_{\rm I-}\rho_-=0,\label{nfb} \\
&& \dot\rho_{\rm R}+4H\rho_{\rm R}-
\mbox{$\sum_{r=\pm}$}\Gamma_{{\rm I}r}\rho_r=0,\label{rR}\\
&& \dot n_{L}+3Hn_{L}-\mbox{$\sum_{r=\pm}$}2\ve_{Lr}\Gamma_{{\rm
I}r}n_r=0,\label{nL}\\
&& \dot n_{\Gr}+3Hn_{\Gr}-C_{\Gr} \lf n^{\rm eq}\rg^2=0.\label{ng}
\end{eqnarray}\eeqs
Here the overdot denotes derivation with respect to the cosmic time 
$t$, $\ve_{Lr}=\sum_i{\Gm[{\rm I}_r\to \nu^c_i]\ve_i/\Gm[{\rm
I}r]}$, and $n_r=\rho_r/m_{{\rm I}r}$. Also, $n^{\rm eq}=
{\zeta(3)T^3/\pi^2}$ is the equilibrium number density of each
bosonic relativistic species, $C_{\Gr}$ is a collision term for
$\Gr$ production which, in the limit of massless MSSM gauginos,
turns out to be \cite{kohri, brand}
\beq C_{\Gr} = \frac{3\pi}{16\zeta(3)\mP^2}\sum_{i=1}^{3} c_i
g_i^2 \ln\left(\frac{k_{i}}{g_i}\right),\eeq
where $(c_i)=(33/5,27,72)$, $g_i$ are the gauge 
coupling constants of the MSSM, and $(k_i)=(1.634,1.312,1.271)$. 
Finally, the Hubble expansion parameter $H$ during this period is 
given by
\begin{equation} \label{Hini}
H=\frac{1}{\sqrt{3}\mP} \left(m_{{\Gr}}n_{{\Gr}}+\rho_-
+\rho_++\rho_{\rm R} \right)^{1/2}.
\end{equation}
Clearly, in the limit of massless MSSM gauginos, the resulting 
$n_{\Gr}$ is practically $m_{\Gr}$-independent. The temperature 
$T$ and the entropy density $s$ are found from the relations
\begin{equation} \rho_{\rm R}=\frac{\pi^2}{30}g_*
T^4~\mbox{and}~s=\frac{2\pi^2}{45}g_* T^3.
\label{rs}\end{equation}

The system of Eqs.~(\ref{nf})-(\ref{ng}) is solved under
the following initial conditions:
\beqs\beq\rho_+(0)=\rho_-(0)=\Vhio/2\eeq and \beq \rho_{\rm
R}(0)=n_{\Gr}(0)=n_{L}(0)=0,\label{init} \eeq\eeqs
where we assumed that the inflationary energy density is 
equally distributed between the oscillatory subsystems 
${\rm I}_+$ and ${\rm I}_-$. This is a reasonable assumption 
since the damped oscillations of I$_+$ and I$_-$ commence 
immediately after the termination of FHI as a consequence of 
the fact that $\msp$ and $\mqn\gg H_{\rm I0}\equiv\sqrt{\Vhio}/
\sqrt{3}\mP$, the inflationary Hubble parameter.

In Fig.~\ref{Trg}, we illustrate the cosmological evolution of the
quantities $\log\rho_+$ (dotted gray line), $\log\rho_-$ (dashed
gray line), $\log\rho_{\rm R}$ (gray line), $\log |Y_L|$
(black solid line), and $\log |Y_{\Gr}|$ (black dashed line)  as
functions of $\log T$ for the values of the parameters given in
the first column of \Tref{tab4} (case A). In particular, these
parameters yield $\msp=2.5\times10^{16}~\GeV$ and
$\Gqp=4.1\times10^{10}~\GeV$ for the ${\rm I}_+$ subsystem, whereas
$\mqn=2.9\times10^{13}~\GeV$ and $\Gqn=0.62~\GeV$ for the 
${\rm I}_-$ subsystem. Since $H_{\rm I0}\simeq1.65\times10^{11}~
\GeV\ll \msp$ and $\mqn$, we verify that the phase of the
oscillations of ${\rm I}_+$ and ${\rm I}_-$ starts immediately 
after the end of FHI. 

\begin{figure}[!t]
\centering\includegraphics[width=60mm,angle=-90]{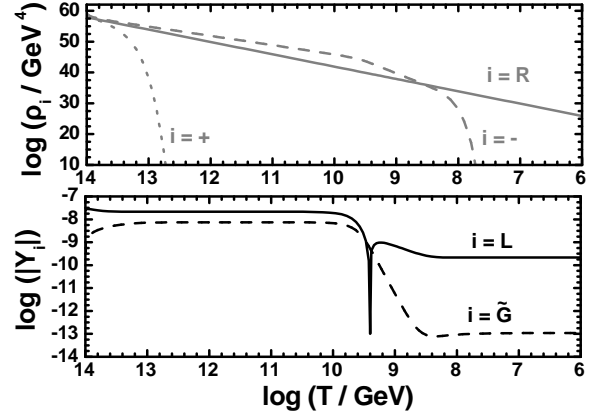}
\caption{\label{Trg} The evolution of the quantities $\log\rho_i$
with $i=+$ (gray dotted line), $i=-$ (gray dashed line), $i={\rm
R}$ (gray line), $\log |Y_L|$ (black solid line), and $\log
|Y_{\Gr}|$ (black dashed line) as functions of $\log T$ for the
values of the parameters in the case A of \Tref{tab4}.}
\end{figure}

From Fig.~\ref{Trg}, we observe that FHI is followed by an extended
matter dominated era, where we have initially the dominance of the 
oscillating and decaying ${\rm I}_+$ and ${\rm I}_-$ subsystems. 
Due to the strong hierarchy between $\Gm[\rm I+]$ and $\Gm[\rm I-]$, 
the decay of ${\rm I}_+$ occurs very early at $T=T_{+}\simeq7.2\times 
10^{13}~\GeV$ -- this temperature corresponds to the intersection 
of the $\rho_+$ and $\rho_{\rm R}$ lines in Fig.~\ref{Trg}. An 
approximate estimate of this temperature can be obtained from 
Eq.~(\ref{Trh}) by replacing $\Gamma_{{\rm I}-}$ with 
$\Gamma_{{\rm I}+}$. This estimate is about $8.8\times10^{13}~\GeV$, 
which is quite close to the value of $T_{+}$ found numerically. 
After the ${\rm I}_+$ decay, the ${\rm I}_-$ subsystem continues 
its oscillations until 
$\rho_-$ meets $\rho_{\rm R}$ at $\Trh=3.5\times10^{8}~\GeV$. This 
numerical result is in excellent agreement with the estimate 
obtained by using Eq.~(\ref{Trh}), which is listed in the column A 
of \Tref{tab4}. After reheating, the universe enters a conventional 
radiation dominated era. Therefore, although our scenario involves 
two oscillatory systems, ${\rm I}_+$ and ${\rm I}_-$, the final 
$\Trh$ can be accurately computed by \Eref{Trh} thanks to the strong 
hierarchy encountered between $\Gamma_{{\rm I}+}$ and 
$\Gamma_{{\rm I}-}$.

In Fig.~\ref{Trg}, we also depict the cosmological evolution of the 
absolute values of the lepton abundance $Y_{L}=n_{L}/s$ and the 
gravitino abundance $Y_{\Gr}=n_{{\Gr}}/s$. We see that $|Y_{L}|$ 
and $|Y_{\Gr}|$, immediately after the decay of the ${\rm I}_+$ 
subsystem, reach constant values equal to $3\times10^{-9}$ and 
$2.6\times10^{-8}$ respectively. However, they are later strongly 
diluted due to the entropy release during the subsequent decay of 
the ${\rm I}_-$ subsystem. The lepton abundance $Y_L$ at $T=T_+$ 
originates from the lepton asymmetry $2\ve_{L+}$ generated by the 
decay of one ${\rm I}_+$ inflaton -- $\ve_{L+}$ is defined just 
below \Eref{ng}. However, the subsequent decay of the ${\rm I}_-$ 
subsystem gives rise to a new lepton asymmetry $2\ve_{L-}$
per decaying inflaton. Note that the sign of this new asymmetry, 
which survive for $T<\Trh$, is opposite to the sign of the earlier 
one which was diluted. As a consequence of this cosmological 
evolution, the present values of both $Y_{L}$ and $Y_{\Gr}$ are 
generated close to $T\simeq\Trh$. Numerically, we find that
$Y_{L}=-2\times10^{-10}$ and $Y_{\Gr}=10^{-13}$, which are in good
agreement with the values obtained by using \eqs{Yb}{Ygr} in the 
case A of \Tref{tab4} -- note that the corresponding $Y_{B}$ turns 
out to be $7.6\times10^{-11}$. Therefore, we see that \eqs{Yb}{Ygr}, 
despite their simplicity, give a very accurate determination of 
$Y_B$ and $Y_{\Gr}$ in our set-up. 

\def\ijmp#1#2#3{{Int. Jour. Mod. Phys.}
{\bf #1},~#3~(#2)}
\def\plb#1#2#3{{Phys. Lett. B }{\bf #1},~#3~(#2)}
\def\zpc#1#2#3{{Z. Phys. C }{\bf #1},~#3~(#2)}
\def\prl#1#2#3{{Phys. Rev. Lett.}
{\bf #1},~#3~(#2)}
\def\rmp#1#2#3{{Rev. Mod. Phys.}
{\bf #1},~#3~(#2)}
\def\prep#1#2#3{{Phys. Rep. }{\bf #1},~#3~(#2)}
\def\prd#1#2#3{{Phys. Rev. D }{\bf #1},~#3~(#2)}
\def\npb#1#2#3{{Nucl. Phys. }{\bf B#1},~#3~(#2)}
\def\npps#1#2#3{{Nucl. Phys. B (Proc. Sup.)}
{\bf #1},~#3~(#2)}
\def\mpl#1#2#3{{Mod. Phys. Lett.}
{\bf #1},~#3~(#2)}
\def\arnps#1#2#3{{Annu. Rev. Nucl. Part. Sci.}
{\bf #1},~#3~(#2)}
\def\sjnp#1#2#3{{Sov. J. Nucl. Phys.}
{\bf #1},~#3~(#2)}
\def\jetp#1#2#3{{JETP Lett. }{\bf #1},~#3~(#2)}
\def\app#1#2#3{{Acta Phys. Polon.}
{\bf #1},~#3~(#2)}
\def\rnc#1#2#3{{Riv. Nuovo Cim.}
{\bf #1},~#3~(#2)}
\def\ap#1#2#3{{Ann. Phys. }{\bf #1},~#3~(#2)}
\def\ptp#1#2#3{{Prog. Theor. Phys.}
{\bf #1},~#3~(#2)}
\def\apjl#1#2#3{{Astrophys. J. Lett.}
{\bf #1},~#3~(#2)}
\def\n#1#2#3{{Nature }{\bf #1},~#3~(#2)}
\def\apj#1#2#3{{Astrophys. J.}
{\bf #1},~#3~(#2)}
\def\anj#1#2#3{{Astron. J. }{\bf #1},~#3~(#2)}
\def\mnras#1#2#3{{MNRAS }{\bf #1},~#3~(#2)}
\def\grg#1#2#3{{Gen. Rel. Grav.}
{\bf #1},~#3~(#2)}
\def\s#1#2#3{{Science }{\bf #1},~#3~(#2)}
\def\baas#1#2#3{{Bull. Am. Astron. Soc.}
{\bf #1},~#3~(#2)}
\def\ibid#1#2#3{{\it ibid. }{\bf #1},~#3~(#2)}
\def\cpc#1#2#3{{Comput. Phys. Commun.}
{\bf #1},~#3~(#2)}
\def\astp#1#2#3{{Astropart. Phys.}
{\bf #1},~#3~(#2)}
\def\epjc#1#2#3{{Eur. Phys. J. C}
{\bf #1},~#3~(#2)}
\def\nima#1#2#3{{Nucl. Instrum. Meth. A}
{\bf #1},~#3~(#2)}
\def\jhep#1#2#3{{J. High Energy Phys.}
{\bf #1},~#3~(#2)}
\def\jcap#1#2#3{{J. Cosmol. Astropart. Phys.}
{\bf #1},~#3~(#2)}
\def\apjs#1#2#3{{Astrophys. J. Suppl.}
{\bf #1},~#3~(#2)}
\def\arxiv#1{{arXiv:}{#1}}

\bibliographystyle{aipprocl}

\end{document}